\documentclass[12pt,preprint]{aastex}
\pdfoutput=1

\usepackage{amsmath}
\usepackage{amsfonts}
\usepackage{dsfont}
\usepackage{amsxtra}
\usepackage{hyperref}
\usepackage{amssymb}

\usepackage{graphicx,epsfig}
\usepackage{dcolumn}
\usepackage{bm}

\newcommand{\be}{\begin{equation}}
\newcommand{\ee}{\end{equation}}
\newcommand{\bea}{\begin{eqnarray}}
\newcommand{\eea}{\end{eqnarray}}
\newcommand{\beaa}{\begin{eqnarray*}}
\newcommand{\eeaa}{\end{eqnarray*}}
\newcommand{\ba}{\begin{array}}
\newcommand{\ea}{\end{array}}
\newcommand{\bi}{\begin{itemize}}
\newcommand{\ei}{\end{itemize}}
\newcommand{\ben}{\begin{enumerate}}
\newcommand{\een}{\end{enumerate}}

\newcommand{\ra}{\rightarrow}
\newcommand{\lra}{\longrightarrow}

\newcommand{\td}{\tilde}

\newcommand{\lb}{\label}
\newcommand{\g}{\gamma}

\newcommand{\al}{\alpha}
\newcommand{\bt}{\beta}

\newcommand{\dl}{\delta}

\newcommand{\sm}{\sigma}

\newcommand{\WMAP}{\textsl{WMAP} }

\begin{document}


\title{ Spectral components analysis of diffuse emission processes
\bigskip
\bigskip
}

\author{Dmitry Malyshev\altaffilmark{1}}



 \affil{
 \bigskip
 Kavli Institute for Particle Astrophysics and Cosmology, \\
Department of Physics and SLAC National Accelerator Laboratory, \\
Stanford University, Stanford, CA 94305, USA
}

\altaffiltext{1}{malyshev at stanford.edu\\
On leave of absence from ITEP, Moscow, Russia, B. Cheremushkinskaya 25}


\begin{abstract}

We develop a novel method to separate the components of a diffuse emission process based on
an association with the energy spectra.
Most of the existing methods 
use some information about the spatial distribution of components,
e.g., closeness to an external template, independence of components etc.,
in order to separate them.
In this paper we propose a method where one puts conditions on the spectra only.
The advantages of our method are:
1) it is {\it internal}: the maps of the components are constructed as combinations
of data in different energy bins,
2) the {\it components may be correlated} among each other,
3) the method is {\it semi-blind}: in many cases, it is sufficient to assume a functional form of the spectra
and determine the parameters from a maximization of a likelihood function.
As an example, we derive the CMB map and the foreground maps
for seven yeas of \WMAP data.
In an Appendix, we present a generalization of the method, where one can also add a number 
of external templates.

\end{abstract}







\section{Introduction}

In many observations, understanding the diffuse emission components plays a significant role
in the interpretation of results.
For instance, diffuse background modeling is crucial for
the observation of the CMB fluctuations in radio frequencies and
for constraining DM annihilation in gamma-rays.

Diffuse emission in a pixel $i$ at an energy $E_\al$ can be described as
a sum over emission components $\mu$
\be
\lb{eq:diff_eq}
d_\al^i = \sum_{\mu=1}^m u_\al^{i \mu} s_\mu^i + r_\al^i,
\ee
where $s_\mu^i$ is the line of sight density of component $\mu$ at pixel $i$,
$u_\al^{i \mu}$ is the corresponding energy spectrum at energy $E_\al$
(the matrices $u$ are also called mixing matrices).
$r_\al^i$ denotes random noise (instrumental or physical).

In various applications it is important to find the energy spectrum and/or
the spatial distribution of the components.
In many cases, one can neglect the dependence of the spectrum on pixels
within the region of interest, i.e. $u_\al^{i \mu} = u_\al^{\mu}$.
Let us also for the moment neglect the noise term $r_\al^i$.
Then the problem is to find $u_\al^{\mu}$ and $s_\mu^i$ from the system of equations
\be
\lb{eq:no-noise}
d_\al^i = \sum_{\mu=1}^m u_\al^{\mu} s_\mu^i.
\ee
Suppose that the total number of energy bins (frequency bands) $k$ is larger
than the total number of significant components of the emission $m$.
We also assume that the number of pixels $N$ is larger than both $k$ and $m$.
Then, in general, this system is both overdetermined and redundant.
It is overdetermined because the rank of the matrix $d^i_\al$ is $k$ while the rank of 
the matrices $u^\mu_\al$ and $s^i_\mu$ is $m$.
It is redundant because there is a symmetry of multiplying the matrix $u$ from the right with
a matrix $A$ and multiplying the matrix $s$ from the left with the inverse of $A$
\be
\lb{eq:degen}
u \lra u\cdot A,\qquad\qquad s \lra A^{-1} \cdot s. 
\ee
Different methods can be classified according to the usage of the 
overdetermination of the system
to most efficiently extract the components of the emission
and how they deal with the redundancy of the system.

One of the most simple cases is when either the spectra $u$ or the spatial distributions $s$
of the components are known.
If $u$ is known then $s$ can be found from Equation (\ref{eq:no-noise}),
and vice versa,
\be
\lb{eq:syst}
\ba{ccl}
s &=& (u^Tu)^{-1} u^T d, \\
u &=& d s^T (s s^T)^{-1}.
\ea
\ee
Denote $f = (u^Tu)^{-1} u^T$. 
The matrix $f$ satisfies $f \cdot u = I$,
it is the Moore-Penrose pseudo-inverse matrix for the rectangular
matrix $u$.
Analogously, $h = s^T (s s^T)^{-1}$ is the pseudo-inverse for $s$, $s\cdot h = I$.

If the uncertainty does not depend on both the energy and the position, ${\sm_\al^i} = \sm$,
then Equations (\ref{eq:syst}) can be also derived by minimizing 
\be
\chi^2 = \sum_{\al, i} \frac{(d_\al^i - u_\al^{\mu} s_\mu^i)^2}{{\sm}^2}.
\ee
There is a summation over the repeated index $\mu$.
In the following we will assume a summation over the repeated indices
unless otherwise stated.

In general, the methods to separate the diffuse emission components 
have the following steps:
\vspace{-1mm}
\ben
\vspace{-3mm}
\item Choose a data analysis principle.
\vspace{-3mm}
\item Find a functional such that the principle from p.1 is realized 
at the minimum (maximum).
\vspace{-3mm}
\item Break the redundancy which is not broken by the functional.
\een
\vspace{-3mm}

In many methods the separating principle is the maximum likelihood.
In the case of Gaussian statistics, 
the minimum of the $\chi^2$ is the maximum likelihood point.
The $\chi^2$ has the change of the basis degeneracy (Equation (\ref{eq:degen})).
Some examples of the ways to break the degeneracy are as follows
(a review of the methods can be found in \cite{2008A&A...491..597L})
\ben
\item
{\bf Independent components analysis} (ICA), \cite{1999ISPL....6..145H, 2002MNRAS.334...53M}.
There is an additional term in the functional that depends on the mutual information
(the minimum corresponds to maximal independence).
\item
{\bf Maximal entropy method} (MEM),
\cite{2007ApJS..170..288H, 1998MNRAS.300....1H,2002MNRAS.336...97S}.
There is an additional term that quantifies a closeness
to some prior (usually an external template).
\item
{\bf Generalized morphological component analysis} (GMCA), \cite{2007ITIP...16.2662B}.
There is an additional term that quantifies the separation of components
in a certain dictionary.
\een
Other methods have a different physical principle to start with.
For example,
\ben
\item {\bf Principal components analysis} (PCA), e.g.,
\cite{1999ASPC..162..363F, 2009MNRAS.395...64S}.
The functional is the variance in the space of data vectors.
The first PCA component corresponds to the direction of the largest variance,
the second component corresponds to the next largest variance direction etc.
The degeneracy is broken by choosing the components in the order of decreasing variance.
\item
{\bf Internal linear combination} (ILC), \cite{2004ApJ...612..633E, 2007ApJS..170..288H}.
One uses the fact that the expected variance of a sum of two components is larger
than the variance of either of the components.
The CMB map is modeled as a linear combination of data vectors at different \WMAP frequencies
with the constraint on the coefficients ensuring that the CMB signal is preserved in the linear combination.
The CMB map is determined by minimizing the variance across the sky of the
linear combination map.
\item
{\bf Correlated component analysis} (CCA), \cite{2005EJASP2005.2400B}.
The basic observables are not the fluxes but the two-point correlators.
One uses some parameterization of the source correlator matrices and the energy spectra of the sources.
The parameters are estimated from the least squared difference between the observed
correlators and the model correlators.
The redundancy is broken by choosing a particular functional form of the spectra
and/or assuming that certain components are uncorrelated.
\een

On general grounds, the degeneracy can be broken by either putting constraints on the 
energy spectra or on the spatial distribution of the components.
Most of the models (apart from the CCA)
assume some properties of the component distributions.
Usually this means either using external templates or assuming some independence of 
the components.
In CCA one uses the functional form of the spectra to break the degeneracy.
This allows one, in principle, to separate correlated components internally.

In this paper, we describe a novel method, which we call the spectral components analysis
(SCA).
In this method, we use the maximum likelihood for the fluxes
and assume a functional form of the energy spectra to break the redundancy.
In comparison with the methods that put constraints on the spatial distribution
of the components,
the new method allows one to study correlated components using only internal data.
In contrast with the CCA method, 
one formulates a model directly for the fluxes rather than for the correlators:
the decomposition is simpler and the calculations are easier in this case.

\section{Spectral components analysis}

In this section we formulate the SCA method to separate the components of 
a diffuse emission process.
The components of the emission are modeled by linear combinations 
of the data in different energy bins.
The degeneracy between the components is broken by assuming 
a functional form of the energy spectra.
The linear combination parameters and the parameters of the spectra
are found from a $\chi^2$ minimization.

To start,
we introduce vectors with components labeled by pixels,
e.g., the data in a bin with the central energy $E_\al$ is written as 
$\vec{d_\al} = \{ d^i_\al \}$ where index $i$ labels the pixels.
In vector notations, Equation (\ref{eq:diff_eq}) takes the form
\be
\vec{d}_\al = \sum_\mu u_\al^{\mu} \vec{s}_\mu + \vec{r}_\al,
\ee
where we assume that the energy spectra $u_\al^{\mu}$ do not depend on pixels.
We assume that the number of the emission components is smaller than the number
of energy bins and
model the spatial distribution of the emission components 
as linear combinations of data vectors
\be
\lb{eq:ILC}
\vec{s}_\mu = f_\mu^\bt \vec{d}_\bt.
\ee
Following the \WMAP terminology \citep{2007ApJS..170..288H},
we will sometimes call the emission components 
determined as linear combinations of the data vectors
by internal linear combination (ILC) vectors.
The spectra $u$ and the linear decomposition coefficients $f$
are found by minimizing the $\chi^2$
\be
\chi^2 = \sum_{\al, i} \frac{(d_\al^i - u_\al^{\mu} f_\mu^\bt {d}_\bt^i)^2}{{\sm_\al^i}^2}.
\ee
The matrices $u$ and $f$ can be determined only up to the change of basis degeneracy,
which corresponds to multiplying $u$ from the right by a matrix $A$
and multiplying $f$ from the left by the inverse matrix $A^{-1}$
\be
\lb{eq:degen}
u \cdot f = u A \cdot A^{-1} f.
\ee
We break the degeneracy by assuming a functional form of the spectra
parameterized by a set of parameters $q$
\be
u_\al^\mu(q) = u^\mu(E_\al; q).
\ee
Before we formulate a general algorithm, let us consider the simplest case.
We fix the parameters of the spectra $q$ and assume homogeneous
independent of energy uncertainty, $\sm$.
Then according to the first equation in (\ref{eq:syst})
\be
f = (u^T u)^{-1} u^T.
\ee
The product of the linear combinations matrix defined by this equation
and the matrix of spectra equals the unit matrix
\be
\lb{eq:lc_cond}
\sum_\al f_\mu^\al u_\al^{\nu} = \dl_\mu^\nu,
\ee
which means that a component with spectrum $u^{\nu}$ contributes only to
the linear combination $f_\nu$ and no other components.
In general, due to different properties of noise in different energy bins,
the conditions in Equation (\ref{eq:lc_cond}) will be slightly violated.

Notice, that $\chi^2$ depends quadratically on $f$ and non-linearly on $q$
\be
\lb{eq:pchi2}
\chi^2(f, q) = \sum_{\al, i} \frac{(d_\al^i - u_\al^{\mu}(q) f_\mu^\bt {d}_\bt^i)^2}{{\sm_\al^i}^2}.
\ee
The algorithm has two steps:
\vspace{-1mm}
\bi
\vspace{-3mm}
\item Find $f$ from quadratic fitting.
\vspace{-3mm}
\item Minimize $\chi^2$ with respect to $q$ by a non-linear fitting.
\ei
\vspace{-3mm}

At each step of the non-linear fitting procedure we choose a set of 
parameters $q$ that describe the spectra and find the best fit $f$ by differentiating 
$\chi^2$ with respect to $f$ and by solving the corresponding linear in $f$ equations
\be
\lb{eq:deff}
\sum_\al u_\al^\mu u_\al^\nu (\vec{d}_\bt \cdot \vec{d}_\g)_{g_\al} f_\nu^\g
= \sum_\al u_\al^\mu (\vec{d}_\al \cdot \vec{d}_\bt)_{g_\al} ,
\ee
where by $(\vec{d}_\bt \cdot \vec{d}_\g)_{g_\al} = g_{\al\, ij} d_\bt^i d_\g^j$
we denote the scalar product in the space with metric $g_\al$.
In our case, it is the inverse of the standard deviation squared
\be
g_{\al\, ij} = \frac{1}{{\sm_\al^i}^2} \dl_{ij}.
\ee
In general, $g_{\al\, ij}$ is the inverse of the covariance matrix.

Equation (\ref{eq:deff}) is a usual linear equation, the non-trivial part
is that the unknowns $f_\nu^\g$ have two indices instead of one.
We can solve this equation, if we choose a single index corresponding to all possible
pairs of indices.
The mathematical structure behind this operation is the tensor product,
i.e., we represent the coefficients
$f_\mu^\bt$ as vectors in a linear space $F$ which is
a tensor product of two linear spaces:
the space $D$ spanned by vectors $\vec{d}_\bt$
and the space $S$ spanned by vectors $\vec{s}_\mu$,
$F = D \otimes S$.
Suppose that the dimensions of spaces $D$ and $S$ are $k$ and $m$ respectively,
then the dimension of $F$ is $k\cdot m$.
The basis vectors in $F$ are denoted as
\be
\vec{f}_\kappa = \vec{d}_\bt \otimes \vec{s}_\mu.
\ee
Loosely speaking the operation of the tensor product is a substitution of a pair of
indices $\{\bt, \mu\}$ by a single index $\kappa$.
Denote $\kappa = \{\bt, \mu\}$ and $\lambda = \{\g, \nu\}$,
then the coefficients $f_\nu^\g$ can be represented as a vector with a single index
$\td{f}_\lambda = f_\nu^\g$.
The right hand side of Equation (\ref{eq:deff}) is a vector
\be
V^{\kappa} = \sum_\al u_\al^\mu (\vec{d}_\al \cdot \vec{d}_\bt)_{g_\al}.
\ee
The sum on the left hand side is a matrix
\be
M^{\kappa\lambda} = \sum_\al u_\al^\mu u_\al^\nu (\vec{d}_\bt \cdot \vec{d}_\g)_{g_\al}.
\ee
In the tensor product notations, Equation (\ref{eq:deff}) takes the form
\be
\sum_\lambda M^{\kappa\lambda} \td{f}_\lambda = V^{\kappa}.
\ee
The solution is
\be
\lb{eq:fsoln}
\td{f}_\lambda = \sum_\kappa {M^{-1}}_{\lambda\kappa} V^{\kappa}.
\ee
This equation gives the minimal value of $\chi^2$ for a given set of parameters $q$.
If we substitute $f_\nu^\g(q) = \td{f}_\lambda(q)$ in Equation (\ref{eq:pchi2}),
we get $\chi^2(q)$ as a function of $q$ only.
Now we can use a non-linear minimization procedure to find the best fit parameters $q_*$.

We note that the spectral components analysis 
allows one to find both the spectra and the spatial distributions
of components with only two rather mild assumptions:
\vspace{-5mm}
\bi
\item
The spatial distribution of components does not depend on the energy.
The only dependence comes in the form of overall normalizations, i.e., energy spectra.
\vspace{-2mm}
\item 
These energy spectra have some functional form.
\ei

\section{Example: \WMAP data}
\lb{sec:example}

In this section we use the SCA method
described in the previous section
to study the \WMAP data \citep{2011ApJS..192...14J}.
The \WMAP experiment measures the fluctuations of temperature
in five frequency bands with the center frequencies (in GHz)
\be
\ba{cccccccccccc}
{\rm K} & {\rm Ka} & {\rm Q} & {\rm V} & {\rm W} \\
22.8 & 33.0 & 40.7 & 60.8 & 93.5
\ea
\ee
The data are naturally split into five vectors corresponding to the
five frequency bands.
In the analysis we use
seven years of data \citep{2011ApJS..192...14J}
with $\sim\! 7'$ pixel size corresponding to
the HEALPix parameter nside = 512
\citep{2005ApJ...622..759G}.
We average the data to get $\sim \! 1^\circ$ pixel size resolution
(nside = 64).
We mask the Galactic plane within $ \pm 30^\circ $,
detected point sources \citep{2011ApJS..192...15G}, 
and the one hundred brightest pixels in the K band which we consider to be outliers.
The remaining number of pixels is 24,228.

In the analysis, instead of using the data itself in the 
formulation of the models in Equation (\ref{eq:ILC})
we use the data vectors randomized with the detector noise.
In this case, the expected $\chi^2$ does not depend on the number of components.
Without randomization, the expected $\chi^2$ decreases as the number of components increases.
For example, in the extreme case of a five-component model for the five frequency bands 
the $\chi^2$ is zero.

We start the analysis by assuming only one component of the emission.
The index of the energy spectrum in this case is $n = -0.13$.
The reduced $\chi^2$ is $\chi^2/dof \approx 14$,
which is relatively large.
We conclude that one component is not sufficient to describe the \WMAP data.

For two components, we get the following indices of the energy spectra
$n_1 = 0.04$ and $n_2 = -3.1$.
The fit in this case is sufficiently good, $\chi^2/dof \approx 1.16$.
Based on the spectra, we conclude that most of the contribution
to these components comes from the CMB and the synchrotron emission respectively.
These are the two strongest components at high latitudes.

Note, that in this approach we don't need to assume that the CMB signal exists.
Instead, we derive from the data that there is a component with an index
$n \approx 0$ which can be interpreted as the CMB fluctuations.

The two-component model is already very good.
If we add a third component and allow the indices to be unconstrained,
then the fit converges to some collinear vectors.
In this case, one may choose to use some of the external knowledge.
In this analysis, we assume that the third significant component is the dust emission
and fix the index of the third component
$n_{\rm dust} = 2$,
which is a usual assumption for the spectrum of the dust emission 
\cite[e.g.,][]{2007ApJS..170..288H}.
Although it is true only approximately \citep{1999ApJ...524..867F},
the exact value is not important since this component is subdominant
to the CMB and the synchrotron emissions at the \WMAP frequencies.
We also fix the index of the CMB emission $n_{\rm CMB} = 0$.
The reduced $\chi^2$ is $\chi^2/dof \approx 0.97$.
The best-fit index of the second component is $n_2 = -2.6$.
We interpret this component as a combination of synchrotron and 
free-free emission.

The components are modeled by taking linear combinations of the \WMAP temperature maps.
The best fit linear combinations are:
\be
\ba{llrrrrrrrrrrrr}
\vec{f}_{\rm CMB} &=& (&\!\!\!\! -0.37 & 0.46 & 0.67 & 0.34 & -0.11\; ) \\
\vec{f}_{\rm syn+ff}  &=& (& 0.84 & -0.18 & -0.42 & -0.28 & 0.04 \; ) \\
\vec{f}_{\rm dust} &=& (& 0.31 & -0.44 & -0.52 & -0.01 & 0.66 \; )
\ea
\ee
Note, that the sum of the components for the CMB vector is approximately equal to $1$ 
\be
\sum_\bt  f_{\rm CMB}^\bt = 0.99.
\ee
In the ILC method \citep{2007ApJS..170..288H}
this is a constraint on the linear combinations describing the CMB map.
In our approach this is a property of the linear combination of data vectors
corresponding to a component with $n = 0$ spectrum.
The sum of components for the foreground models
is less than $0.01$, i.e., the CMB signal is canceled in these models.

The \WMAP temperature maps and the residuals after the three-component model subtraction
are shown in Figures \ref{fig:resid1} and \ref{fig:resid2}.
The residuals are sufficiently small with little large scale structures.
The SCA model of the CMB map is presented in Figure \ref{fig:CMB}.
It is also compared to the \WMAP internal linear combination (ILC)
model of the CMB map \citep{2011ApJS..192...15G}.
In Figure \ref{fig:synch} we present the map of the component corresponding
to the spectral index $n = -2.6$.
We compare it to the sum of the synchrotron and the free-free emission 
\WMAP Markov chain Monte Carlo (MCMC) models \citep{2011ApJS..192...15G}.
In Figure \ref{fig:dust} we compare the SCA dust model to the \WMAP MCMC dust model
\citep{2011ApJS..192...15G}.
Although the SCA models and the WMAP models of the emission components
are derived using completely different methods,
there is a fairly good agreement among the models.

In Figure \ref{fig:cldiff} we compare the angular power spectra of the SCA models
to the corresponding angular power spectra of the \WMAP MCMC models.
The CMB angular power spectra are very similar to each other,
while the spectra for the SCA models of the Galactic emission
have more power at small $\ell$ relative to the \WMAP MCMC models.
In Figure \ref{fig:rms} we present the root mean squared (RMS) of the temperature maps for the
three SCA components at the \WMAP frequencies.

The SCA models generally reproduce the \WMAP emission models.
The SCA models have a higher level of random noise, which is expected
for the models derived from internal data in comparison with the models that also use external data.
The only assumptions that were used to derive the SCA maps of the components
are the power-law energy spectra with fixed indices for the CMB and the dust emission,
while the index for the third component is derived by minimizing the $\chi^2$.

\section{Conclusions}

In the paper we propose a novel method to separate the components of 
a diffuse emission process
based on an association with energy spectra.
The new method is
\vspace{-5mm}
\bi
\item internal (the components are modeled as combinations of data at different energies);
\vspace{-3mm}
\item semi-blind (we assume the functional form of the spectra and fit for the parameters);
\vspace{-3mm}
\item can separate correlated components.
\ei
\vspace{-3mm}
%
Using only internal data we avoid systematic uncertainties
related to the usage of external data.
For instance, in various diffuse emission models one needs to know the 
distributions of atomic, molecular, and ionized gas.
The distribution of the atomic hydrogen HI (${\rm H}_1$) is traced by the hyperfine splitting
emission at 21 cm \citep{2005A&A...440..775K, 2009ARA&A..47...27K}.
The corresponding observations have uncertainties related to the determination of the 
radial velocity of the gas and the 
spin temperature $T_S$ \citep{2009ApJ...693.1250D, 2010arXiv1002.0081J}.
The distribution of the molecular ${\rm H}_2$ gas is traced by 
the emission from the CO molecules 
\citep{2001ApJ...547..792D},
some uncertainties may arise from 
a space dependent conversion coefficient ${\rm X_{CO}}$ 
\citep[e.g.,][]{2004A&A...422L..47S}.
The distribution of ionized hydrogen HII can be modeled 
by studying the H-alpha emission
\citep[e.g.,][]{2003ApJS..146..407F,2008PASA...25..184G}.
In addition to uncertainties in the distribution of the interstellar gas,
the models of cosmic ray (CR) propagation in the Galaxy
\citep{Ginzburg1964, 2007ARNPS..57..285S, 2011CoPhC.182.1156V, 2011arXiv1106.5073C}
have uncertainties in the distribution of the CR sources,
in the propagation parameters, and
in the distribution of the interstellar radiation
\citep[e.g.,][]{2007NuPhS.173...44M}.
In general, the uncertainties arise from the fact that one either
needs to make some assumptions when using 
a certain map as a template for an emission process
or there is some processing involved in constructing 
an emission component from external data.

The advantage of internal methods is based on the following simple 
observation: if a physical process contributes significantly to the data,
then one should be able to use the data to describe the spatial
distribution and the energy spectrum of this emission process.
Otherwise, if the process is insignificant for a particular observation,
then one does not need to include it in the consideration.

In the SCA method, one does not need to assume a priori either the number 
of the emission components, their spectra, or the spatial distributions.
All this information is obtained by fitting the model to the data.
As a result, the SCA method is a unique tool to search for new
components of emission with spectra sufficiently different
from the astrophysical foregrounds.
In template fitting and in galactic propagation frameworks,
a new component can be observed as a residual.
The spectrum and the significance of a ``residual'' component are usually biased.
An advantage of the SCA method is that
the new component and the galactic components are fitted 
to the data simultaneously, i.e. all components in this approach are on equal 
footing and the there is no reason to expect a bias for any of the components.

Another advantage of the SCA method is the possibility to separate
correlated components.
This property distinguishes the SCA method from other
internal data analysis methods such as internal linear combinations,
principal components analysis, or independent components analysis.
One of the disadvantages of the SCA method is that we need to assume a homogeneous
scaling of the components, i.e., that the spectra do not depend on the position in the sky.
Although it is not true in general, it is usually a good first order approximation,
especially if the region of interest is small.

In the paper, we use the SCA method to study seven years of the \WMAP data.
We find that at high latitudes a three component model
describes the data with a sufficiently good accuracy.
We interpret the three components as the CMB, the thermal dust, and a combination
of the synchrotron and the free-free emission.
The derived spatial distributions of the components are generally consistent
with the corresponding spatial distributions provided by the \WMAP collaboration.

The SCA method is sufficiently universal. 
It can be applied to other types of diffuse emission data, 
such as infrared, optical, x-ray, and gamma-ray.
One can also use
any representation allowing linear space interpretation,
e.g., coordinates space, spherical harmonics, wavelets.
The method has a straightforward generalization to the case
where some of the components are modeled by external templates
(Appendix \ref{sec:gen_case}).

{\large \bf Acknowledgments.}

\noindent
The author is thankful to 
Elliott Bloom,
Anna Franckowiak,
Daniel Grin,
David W. Hogg,
Igor Moskalenko, 
Dmitry Prokhorov,
Kendrick Smith,
David Spergel
for stimulating discussions.
This work was supported in part by the US Department of Energy contract to SLAC
no. DE-AC02-76SF0051.
The data analysis have been
done using the HEALPix package \citep{2005ApJ...622..759G}.

\appendix
\section{General case}
\lb{sec:gen_case}

In this appendix we present an algorithm that
includes some number of external templates
together with components modeled as internal linear combinations of data vectors
(ILC vectors).
In the most general form,
we will not put any constraints on the spectra
for some of the ILC vectors.
As we have discussed in the introduction,
if there is more than one ILC vector for which we do not put
constraints on the spectra, then one needs to break the change of the basis
degeneracy (Equation (\ref{eq:degen})).
In the algorithm described below we break this degeneracy
by choosing some sub-matrix of the linear combinations matrix $f$ (Equation (\ref{eq:ILC}))
to be the unit matrix.
Components for which we do not put constraints on the spectra
have an independent scale factor in every energy bin.
The process of subtracting these components is called marginalization.

In the general SCA algorithm we can consider the following models for the components
together with the notations (bold font letters denote vectors, while usual font letters denote
numbers):
\ben
\item
Components with known spatial distributions (templates) 
that we marginalize over, ${\td w}_\al^\nu \bf{\td e}_\nu$ (no summation).
\item
Components with known spatial distribution for which we assume a parametric form of the spectra,
$w(p)_\al^\nu \bf{e}_\nu$ (no summation).
\item
Components with unknown spatial distributions (modeled as ILC vectors) that we marginalize over, 
${\td u}_\al^\mu {\td f}_\mu^\bt \bf{v}_\bt$ (no summation over $\mu$).
\item
Components with unknown spatial distributions (modeled as ILC vectors) 
for which we assume a parametric form of the spectra,
$u_\al^\mu(q) f_\mu^\bt \bf{v}_\bt$ (no summation over $\mu$).
\een
The general $\chi^2$ has the form
\be
\chi^2 = \sum_\al  
\left|{\bf{d}}_\al - u_\al^\mu(q) f_\mu^\bt {\bf{v}}_\bt
- {\td u}_\al^\mu {\td f}_\mu^\bt {\bf{v}}_\bt
-w(p)_\al^\nu {\bf{e}}_\nu -{\td w}_\al^\nu {\bf{\td e}}_\nu
\right|_{g_\al}^2,
\ee
where $|{\bf{v}}|_{g_\al}^2 \equiv ({\bf{v}}, {\bf{v}})_{g_\al}$ 
is the norm in space with metric ${g_\al}$.
The metric is given by the inverse of the standard deviation squared,
or, in general, by the inverse of the covariance matrix.

The algorithm has the following steps:
\vspace{-3mm}
\ben
\item Marginalize over ${\td w}_\al^\nu$.
This is equivalent to projecting the vectors $\bf{d}_\al$,
$\bf{v}_\bt$ and $\bf{e}_\nu$ onto the space perpendicular to the space spanned by
 $\bf{\td e}_\nu$.
The residual $\chi^2$ has the form
\be
\chi^2 = \sum_\al  
\left|{\bf{d}}_\al^\perp - u_\al^\mu(q) f_\mu^\bt {\bf{v}}_\bt^\perp
- {\td u}_\al^\mu {\td f}_\mu^\bt {\bf{v}}_\bt^\perp
-w(p)_\al^\nu {\bf{e}}_\nu^\perp
\right|_{g_\al}^2.
\ee
\item
Choose parameters $p$.
Subtract $w(p)_\al^\nu \bf{e}_\nu^\perp$ from $\bf{d}_\al^\perp$ and $\bf{v}_\al^\perp$.
\item
Choose ${\td f}_\mu^\bt$ and marginalize over the parameters
${\td u}_\al^\mu$ by
projecting $\bf{d}_\al^\perp$ on $ {\td f}_\mu^\bt \bf{v}_\bt^\perp$. 
\item
Choose parameters $q$ and marginalize over $f_\mu^\bt$
(Equations (\ref{eq:pchi2}) to (\ref{eq:fsoln})).
\item 
Repeat steps 2, 3, 4, (and 5) to find the best fit non-linear
parameters $p_*$, $q_*$, and ${\td f}_*$.
\een

Given the best fit parameters $p_*$, $q_*$, and ${\td f}_*$,
the maps of the components are reconstructed as follows:
\vspace{-3mm}
\ben
\item
Components with templates and constrained energy spectra are given by
$w(p_*)_\al^\nu \bf{e}_\nu$ (no summation).
In order to get the other maps, we subtract these components from the data
and define ${\bf{d}}'_\al = {\bf{d}}_\al - w(p_*)_\al^\nu {\bf{e}}_\nu$.
\item
The maps for the templates with marginal spectra are obtained
by marginalization of ${\bf{d}}'_\al $ with respect to the templates $\bf{\td e}_\nu$.
The scaling coefficients are obtained from minimizing the $\chi^2$
\be
{\td w}_\al^\nu = \sum_\mu ({\bf{d}}'_\al, {\bf{\td e}}_\mu)_{g_\al} E^{-1\,\mu\nu},
\ee
where $E^{-1\,\mu\nu}$ is the inverse of the matrix of scalar products
$E_{\mu\nu} = ({\bf{\td e}}_\mu, {\bf{\td e}}_\nu)_{g_\al}$.
The maps are given by ${\td w}_\al^\nu \bf{\td e}_\nu$ (no summation).
\item
The ILC components are obtained by linear combinations of model vectors ${\bf{v}'_\al}$
derived from the data vectors ${\bf{d}'_\al}$. If ${\bf{v}_\al} = {\bf{d}_\al}$, then 
${\bf{v}'_\al} = {\bf{v}_\al} - w(p_*)_\al^\nu {\bf{e}}_\nu$.
We also marginalize over the templates ${\bf{\td e}}_\nu$ with unconstrained spectra
by projecting ${\bf{d}'_\al}$ and ${\bf{v}'_\al}$ on ${\bf{\td e}}_\nu$:
${\bf{d}'_\al} \ra {\bf{d}'_\al}^{\perp}$ and ${\bf{v}'_\al} \ra {\bf{v}'_\al}^{\perp}$.
The maps for the ILC components with marginal spectra are given by
${\td u}_\al^\mu {\td f}_{ * \mu}^\bt {\bf{v}'}^{\perp}_{\bt}$ (no summation over $\mu$).
The overall coefficients ${\td u}_\al^\mu$ are found similarly to ${\td w}_\al^\nu$
by minimizing the residual $\chi^2$
that depends on ${\bf{d}'_\al}^{\perp}$
\be
{\td u}_\al^\nu = \sum_\mu ({\bf{d}'}^{\perp}_\al, {\bf{\td f}}_\mu)_{g_\al} F^{-1\,\mu\nu},
\ee
where $ {\bf{\td f}}_\mu = {\td f}_{ * \mu}^\bt {\bf {v'}}^{\perp}_{\bt}$
are the ILC combinations
and $F^{-1\,\mu\nu}$ is the inverse of the matrix of scalar products
$F_{\mu\nu} = ( {\bf{\td f}}_\mu,  {\bf{\td f}}_\nu)_{g_\al}$.
\item
The maps for the ILC components with functional forms of the spectra are equal to
$u_\al^\mu(q_*) f_\mu^\bt {\bf{v'}}^{\perp}_{\bt}$ (no summation over $\mu$).
The coefficients $f_\mu^\bt$ are found from Equations (\ref{eq:deff}) to (\ref{eq:fsoln}),
where we substitute $u_\al^\mu \ra u_\al^\mu(q_*)$ and ${\bf d}_\al \ra {\bf{d'}}^{\perp}_\al$.
\een
The general SCA method gives simultaneously the spectra and the maps
of all four types of models for the emission components:
templates with and without assuming a functional form of the spectra
and ILC combinations 
with and without assuming a functional form of the spectra.

We note that in some cases instead of a non-linear fitting procedure one
can use a convergent iterative process 
\citep[e.g.][]{2007ITIP...16.2662B, 2012arXiv1201.3370T}
to define the spectra and the spatial distributions of the emission components.

\newcommand{\onepic}{0.5}
\newcommand{\twopic}{0.42}
\newcommand{\threepic}{0.45}

\begin{figure}[t] 
\vspace{-5mm}
\begin{center}
\hspace{-10mm}
\epsfig{figure = 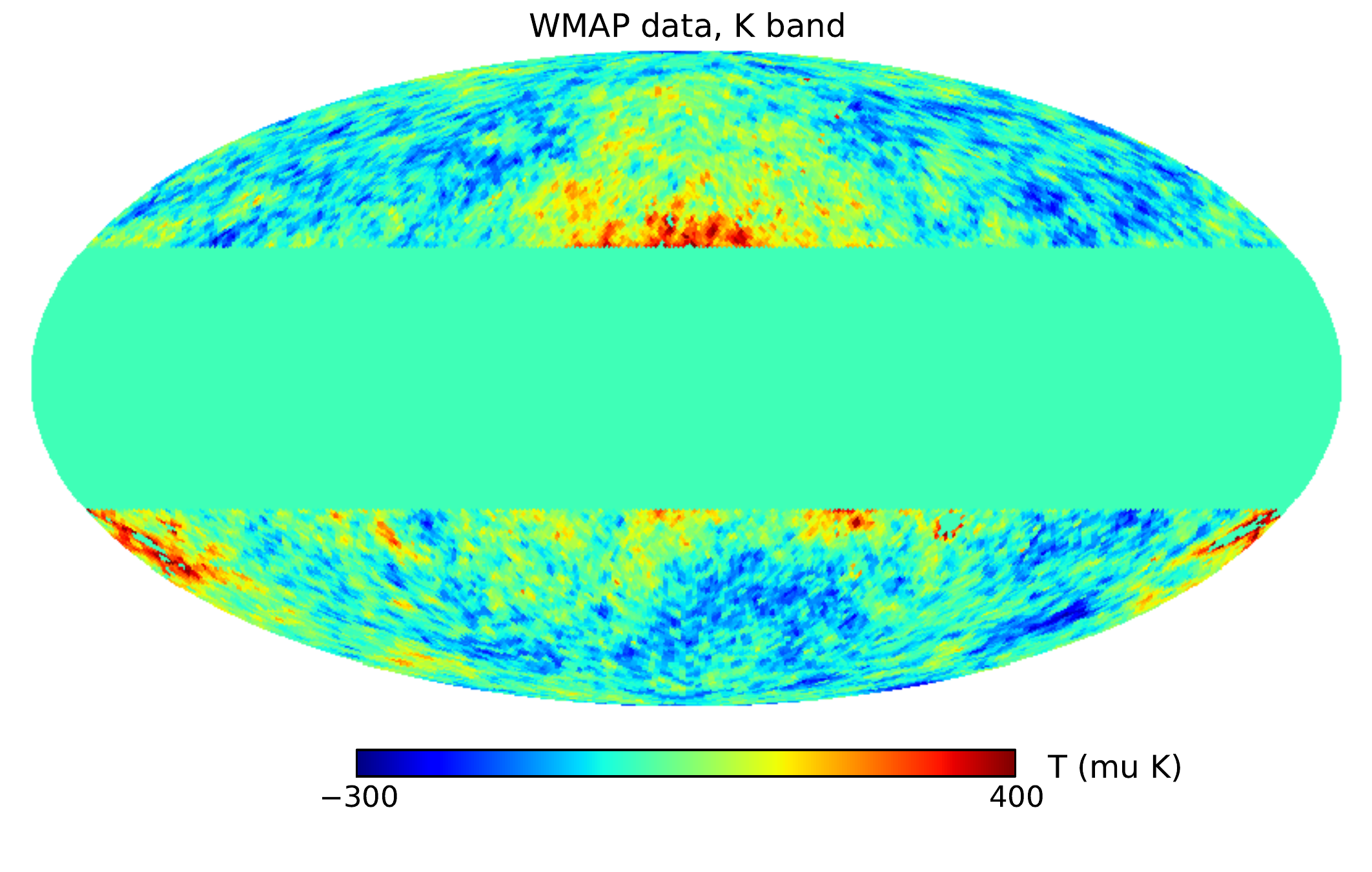, scale=\twopic}
\epsfig{figure = 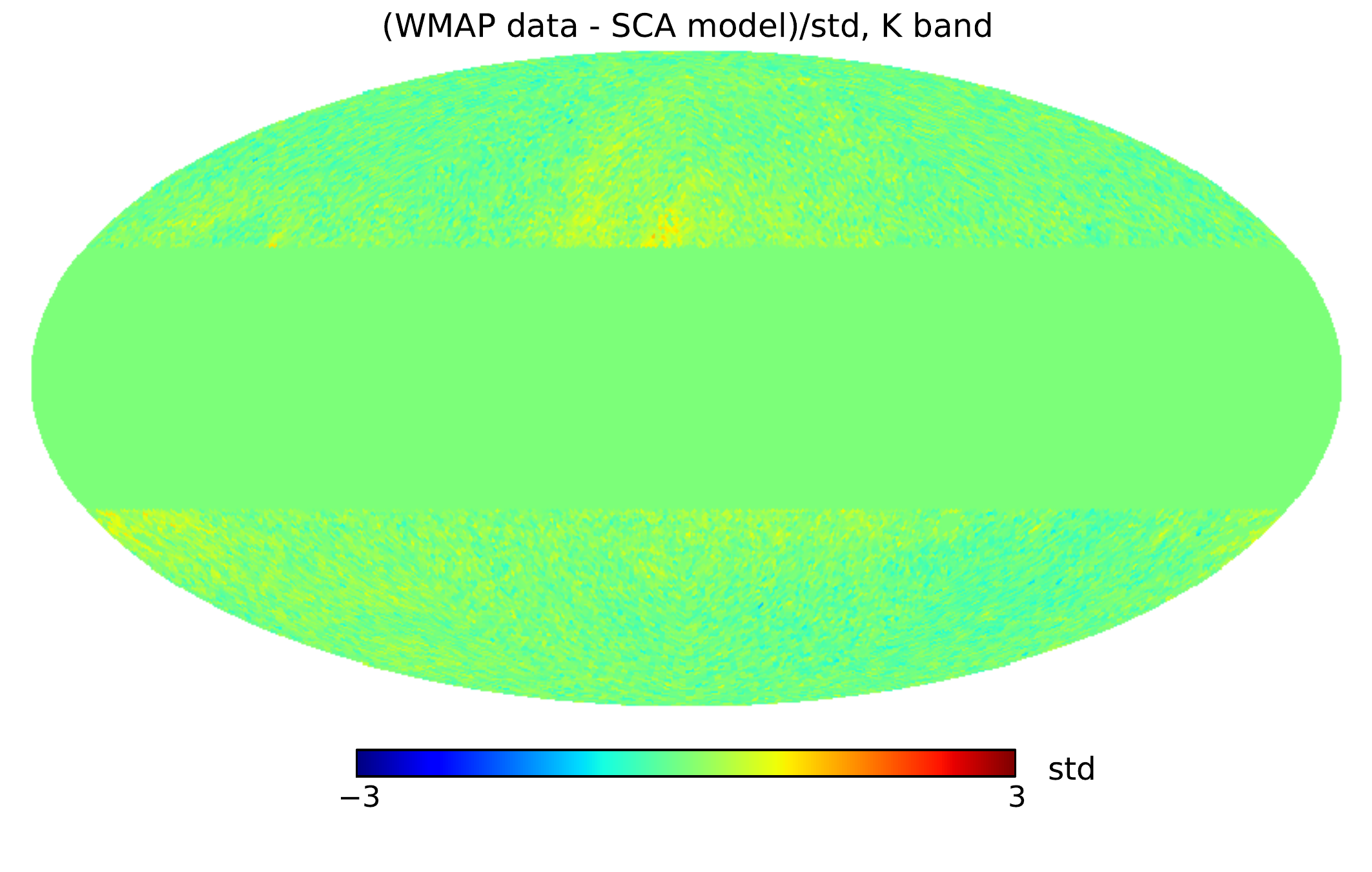, scale=\twopic} \\
\hspace{-10mm}
\epsfig{figure = 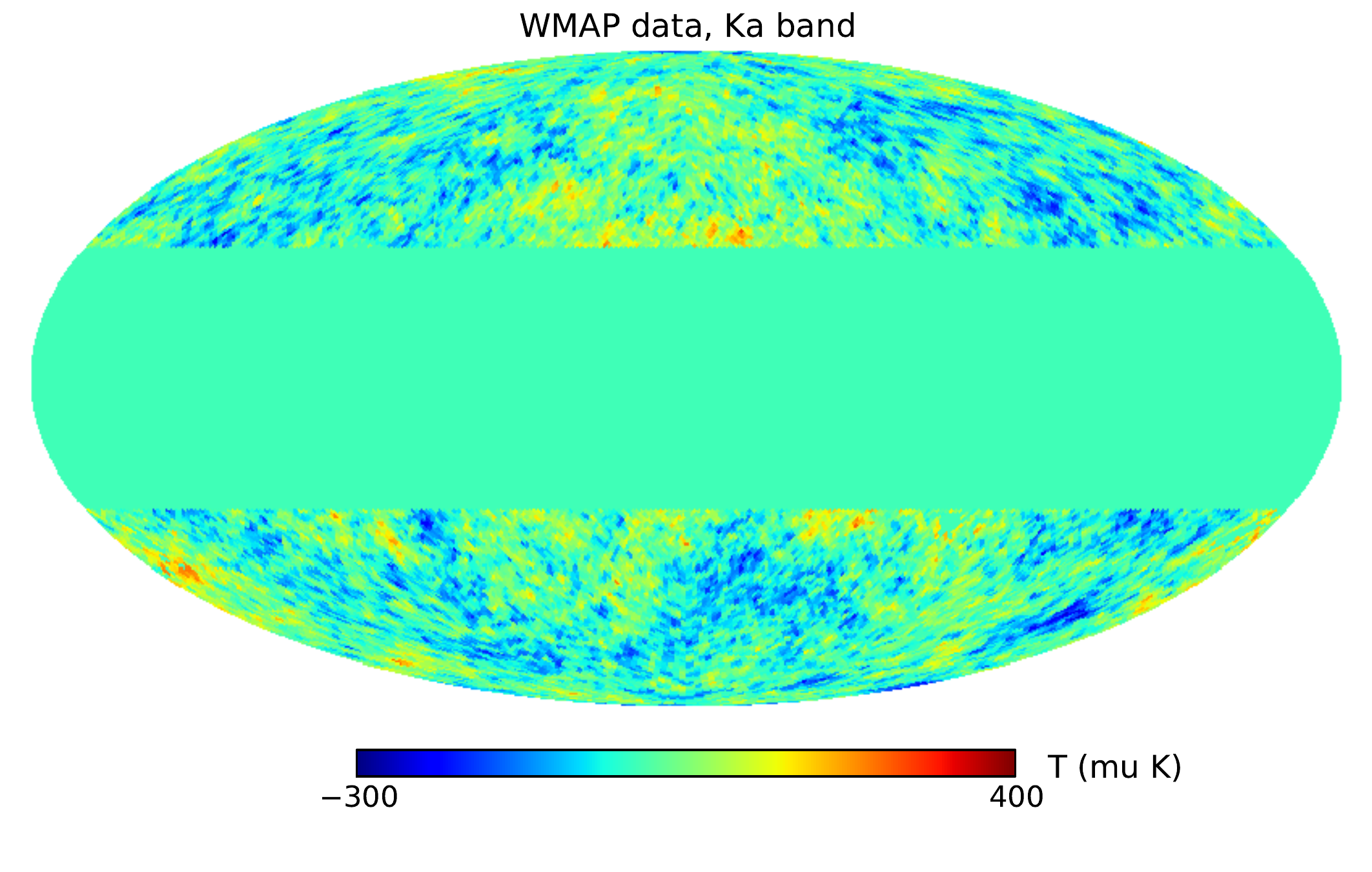, scale=\twopic}
\epsfig{figure = 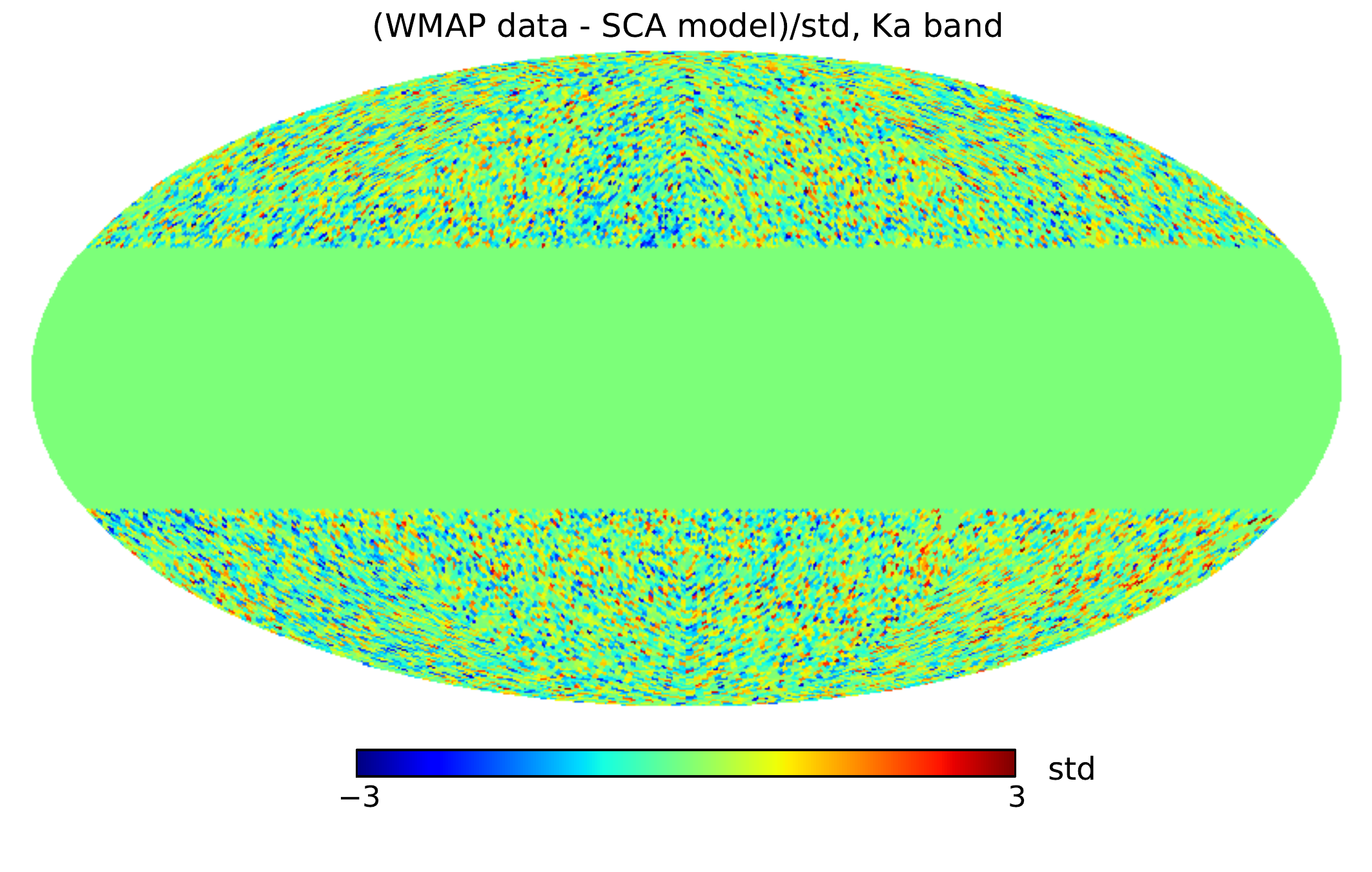, scale=\twopic} \\
\hspace{-10mm}
\epsfig{figure = 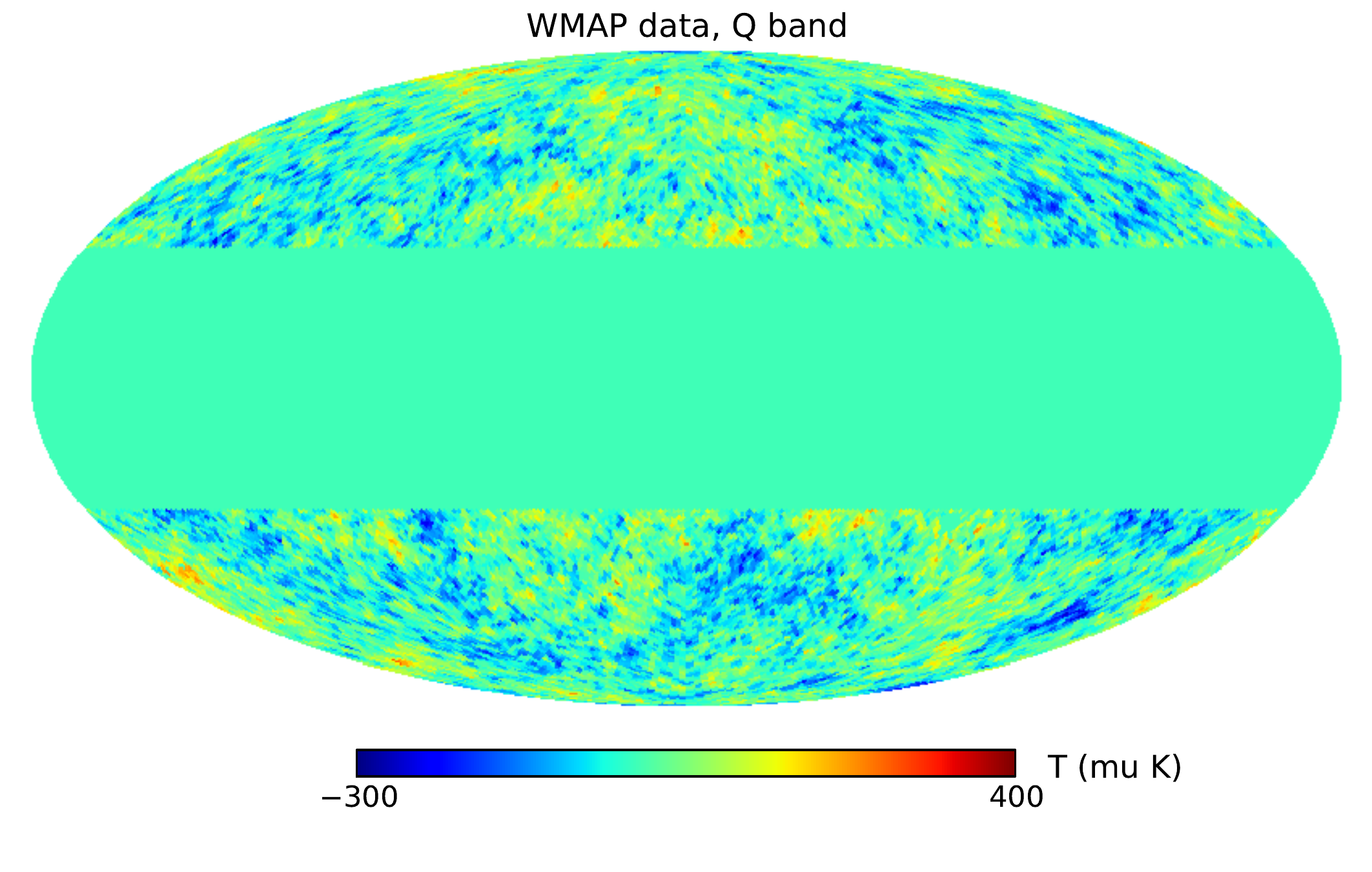, scale=\twopic}
\epsfig{figure = 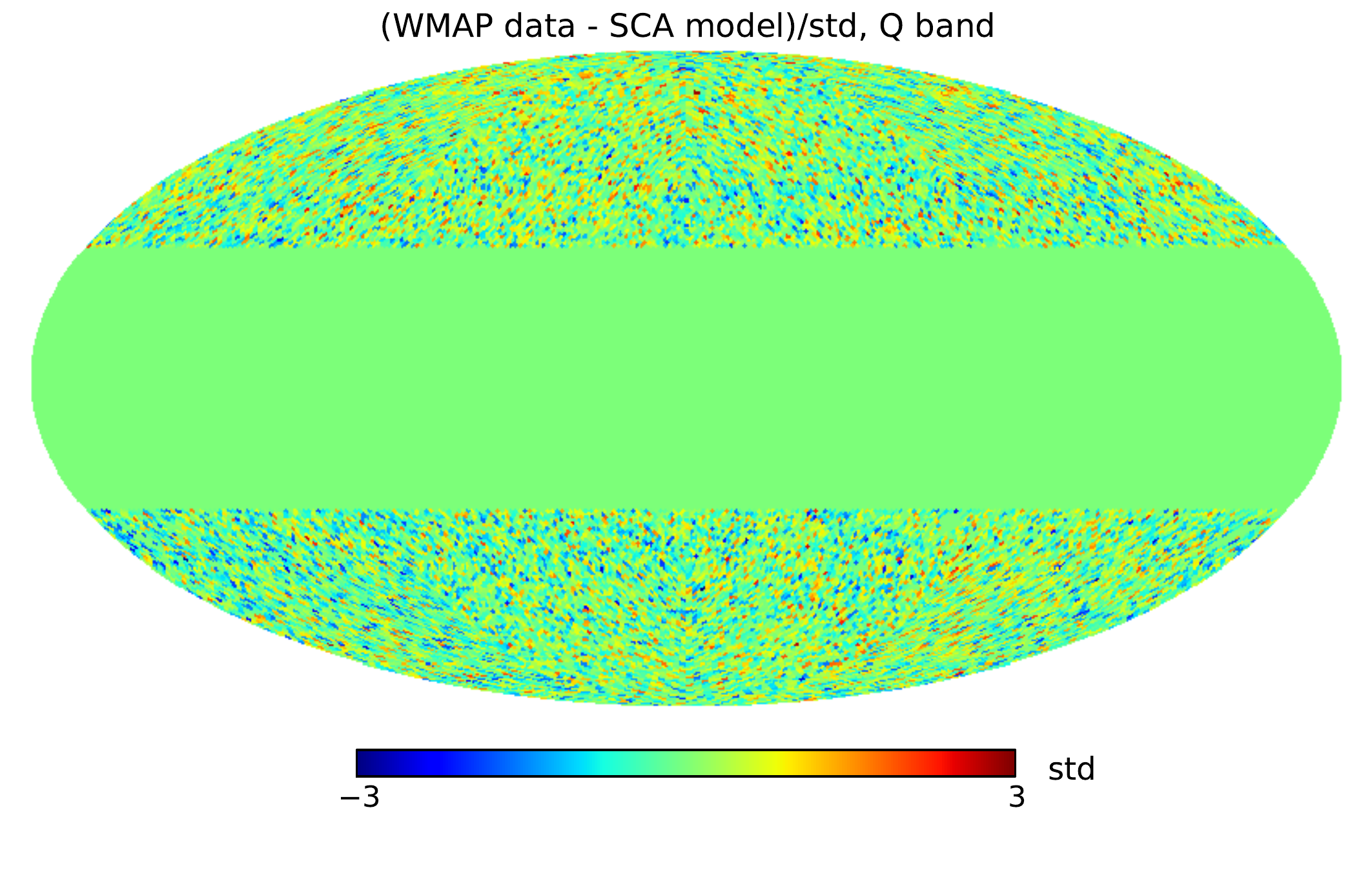, scale=\twopic} 
\end{center}
\vspace{-8mm}
\noindent
\caption{\small 
Left: maps of seven years of \WMAP data 
in K, Ka, and Q frequency bands \citep{2011ApJS..192...14J}.
Right: residuals after subtracting the SCA models (Figures
\ref{fig:CMB}, \ref{fig:synch}, and \ref{fig:dust})
divided by the instrumental noise.
}
\label{fig:resid1}
\vspace{1mm}
\end{figure}

\begin{figure}[t] 
\vspace{-5mm}
\begin{center}
\hspace{-10mm}
\epsfig{figure = 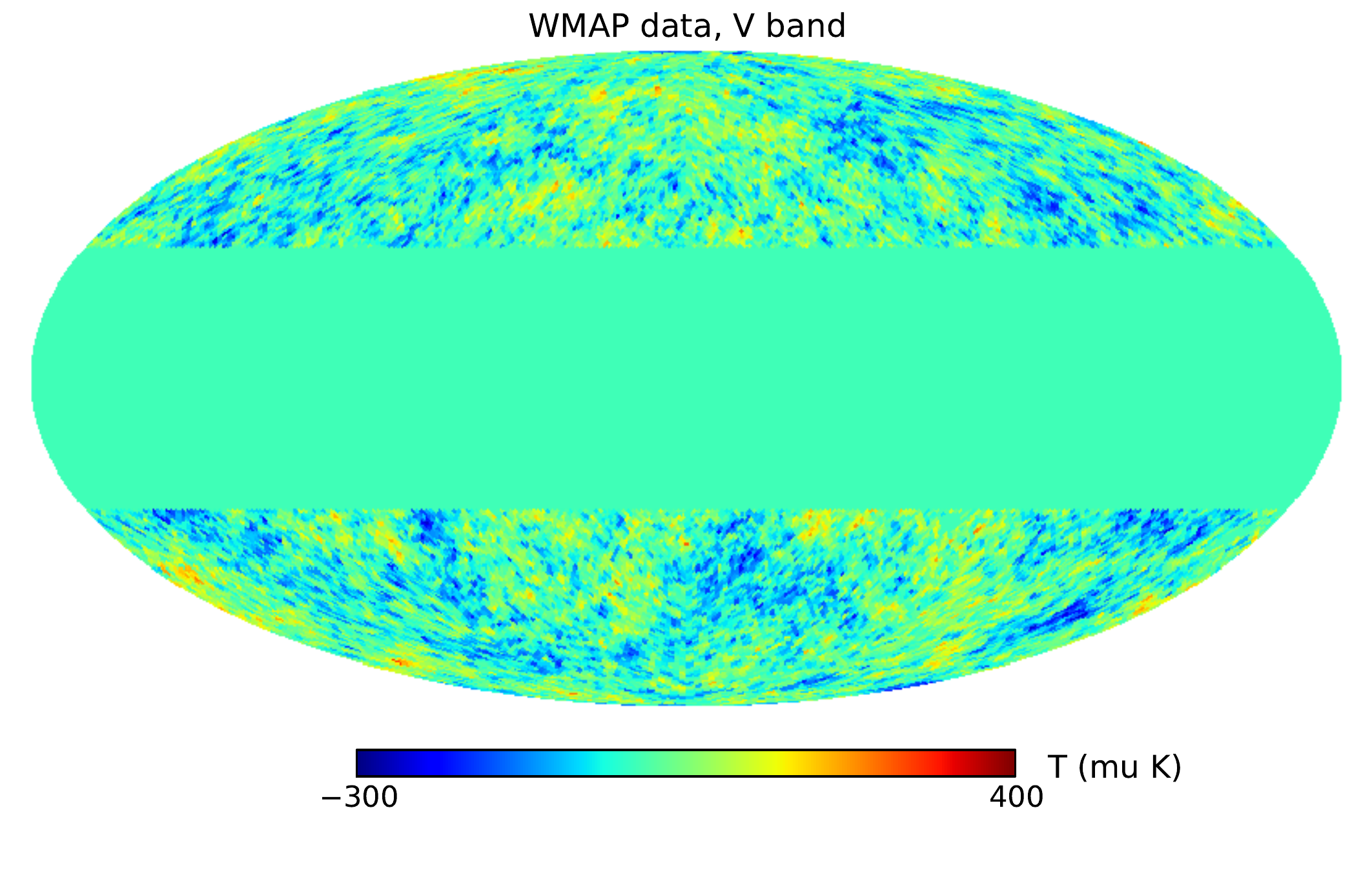, scale=\twopic}
\epsfig{figure = 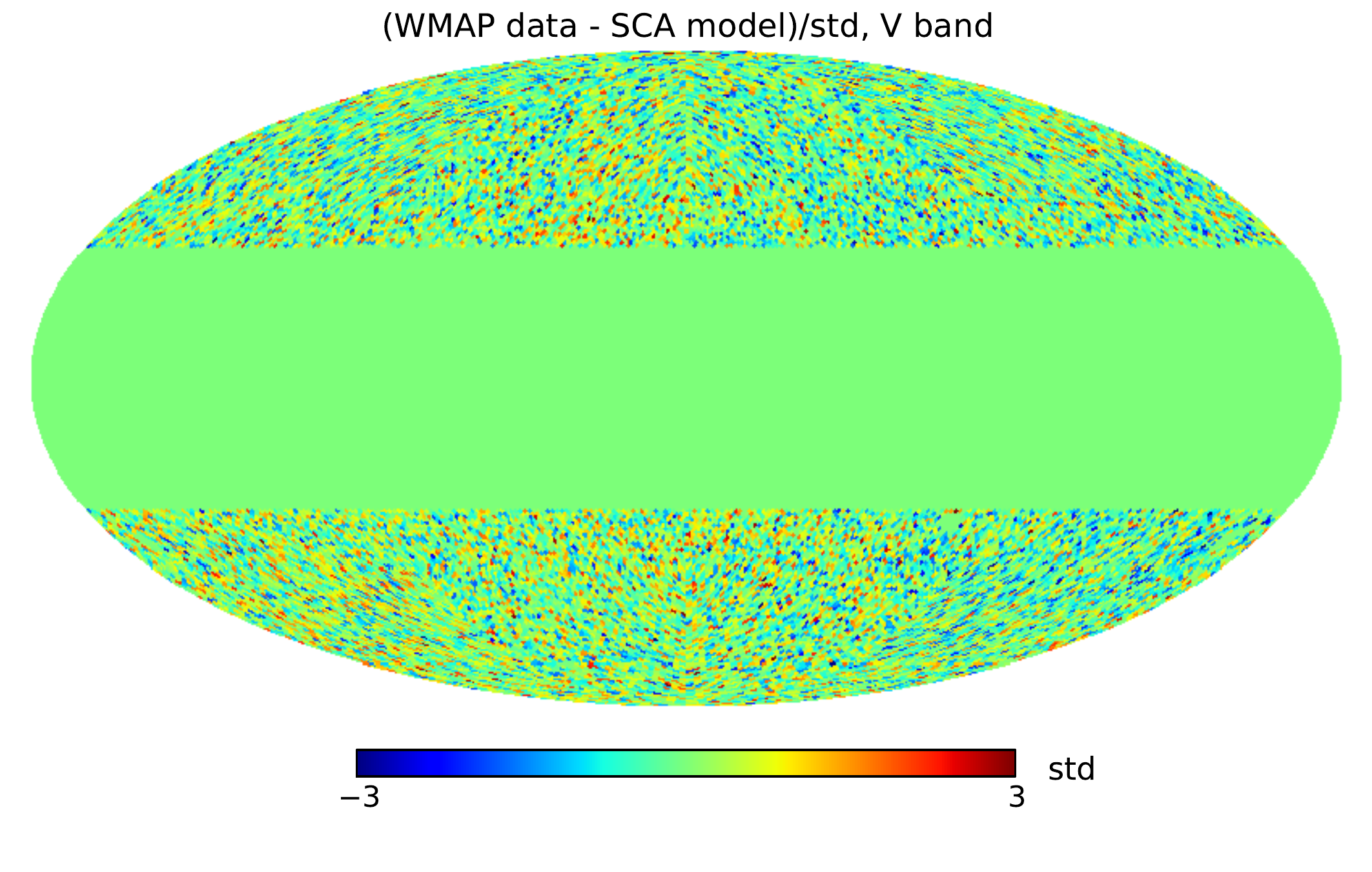, scale=\twopic} \\
\hspace{-10mm}
\epsfig{figure = 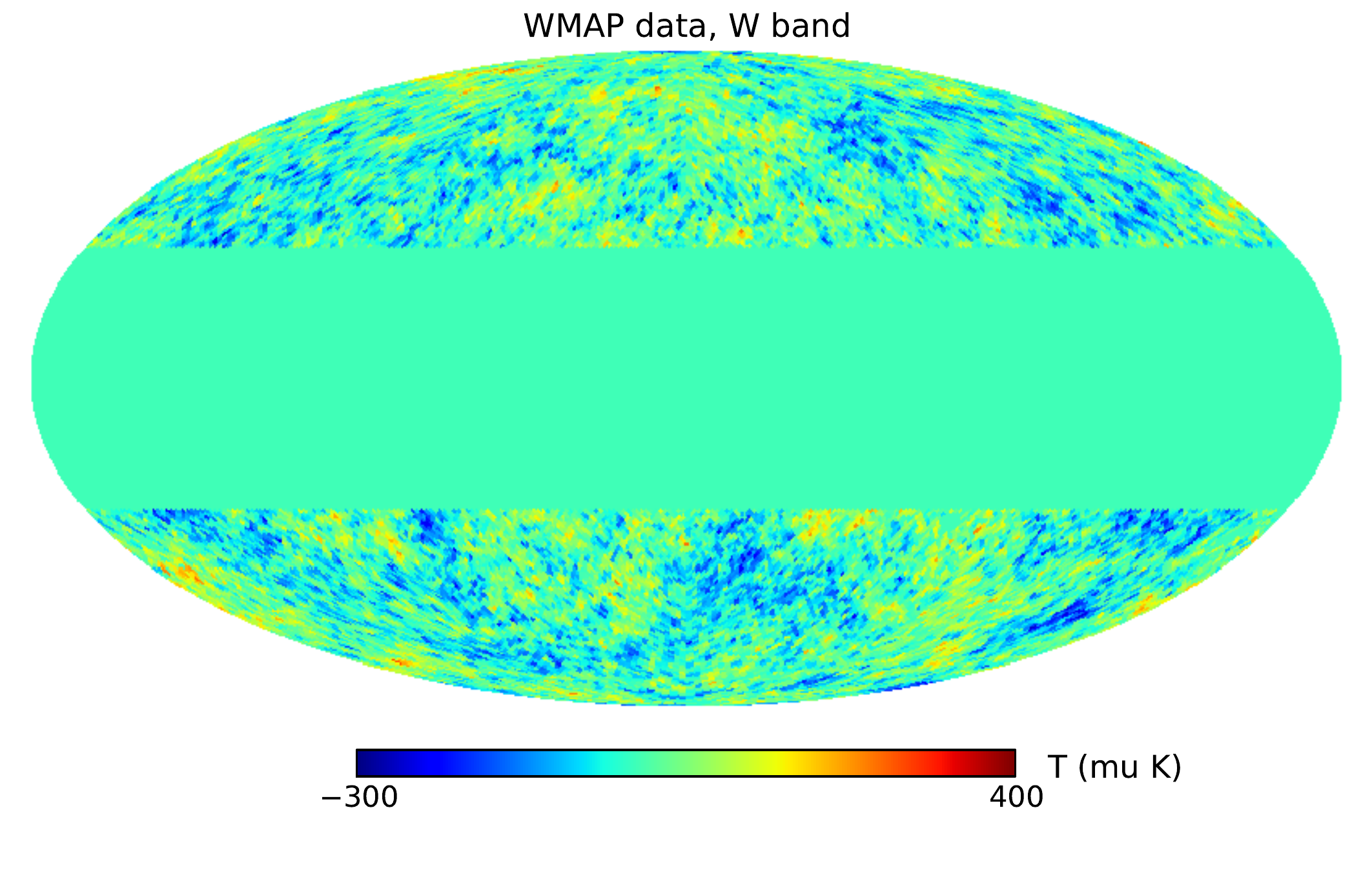, scale=\twopic}
\epsfig{figure = 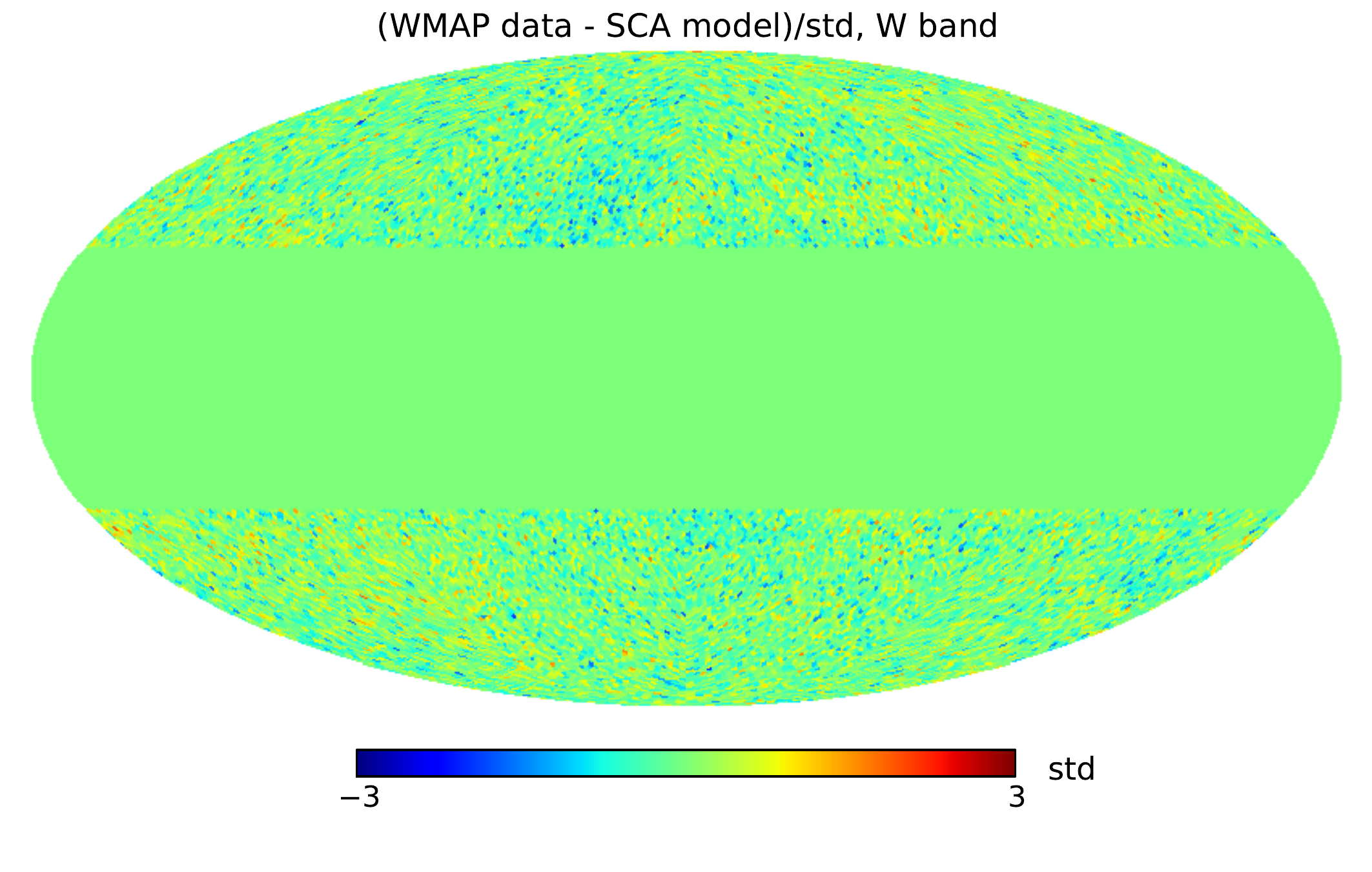, scale=\twopic} 
\end{center}
\vspace{-8mm}
\noindent
\caption{\small
Same as in Figure \ref{fig:resid1} for V and W frequency bands.}
\label{fig:resid2}
\vspace{1mm}
\end{figure}

\begin{figure}[t] 
\vspace{-5mm}
\begin{center}
\epsfig{figure = 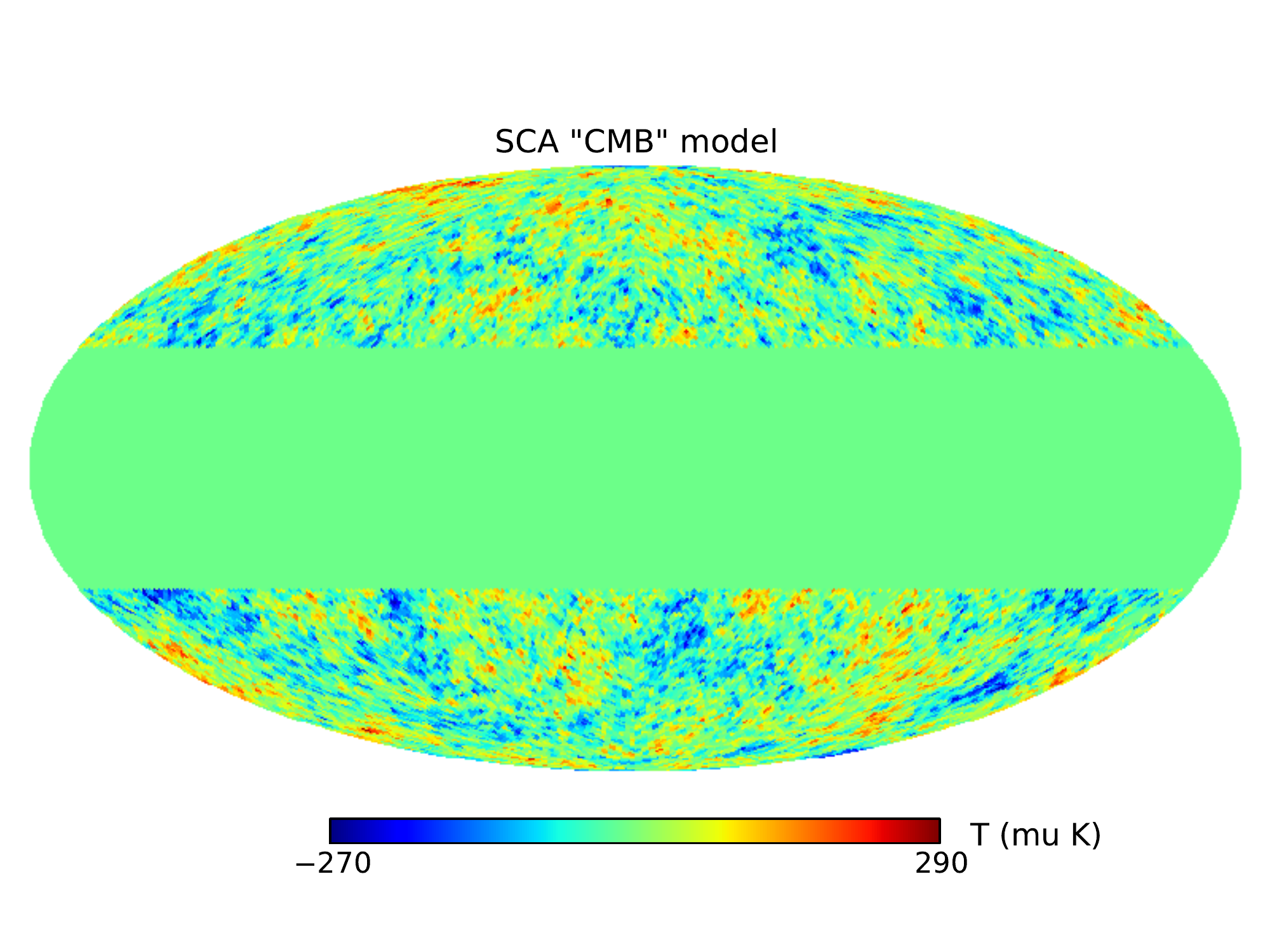, scale=\threepic} \\
\vspace{-5mm}
\epsfig{figure = 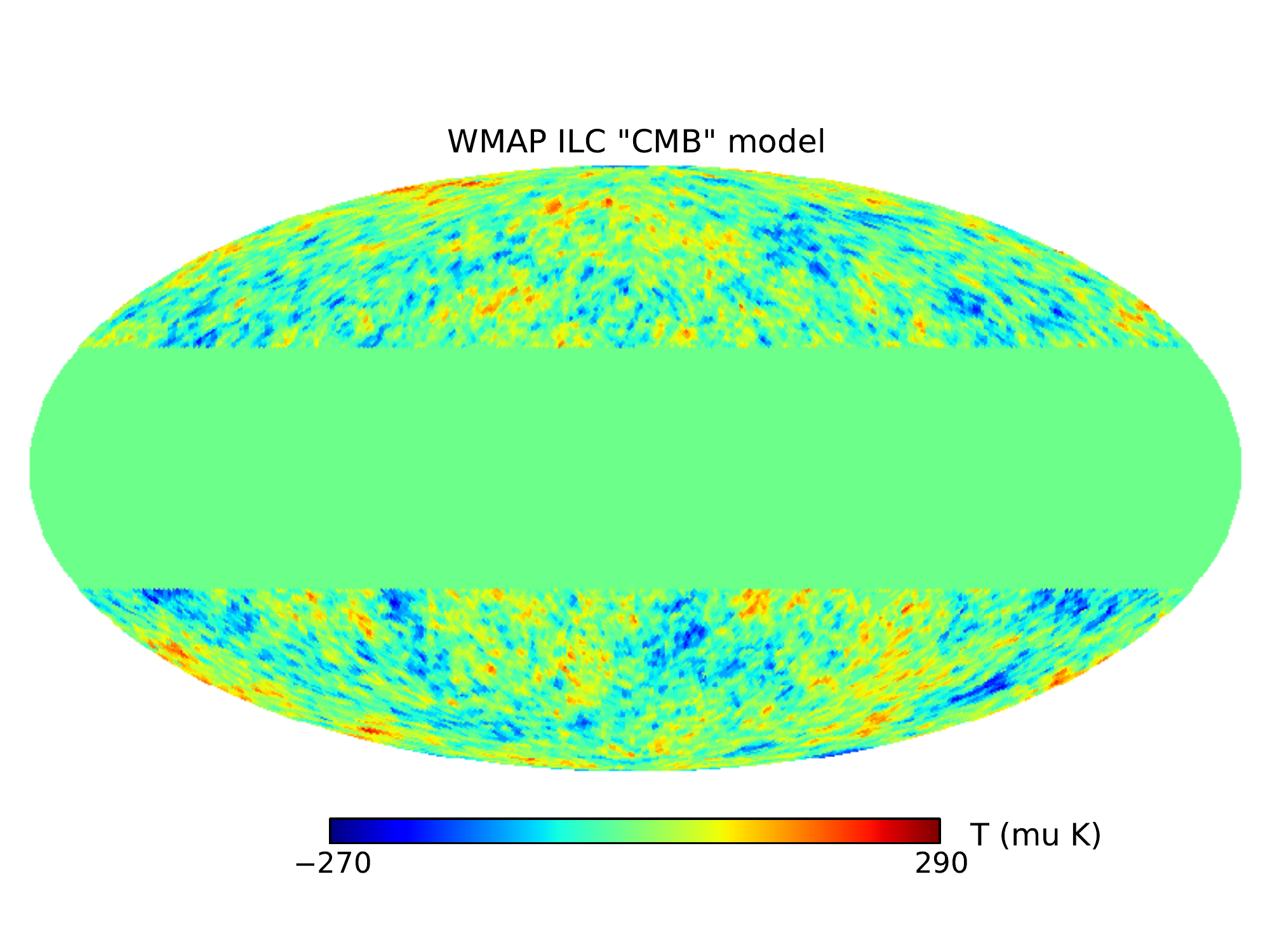, scale=\threepic} \\
\vspace{-5mm}
\epsfig{figure = 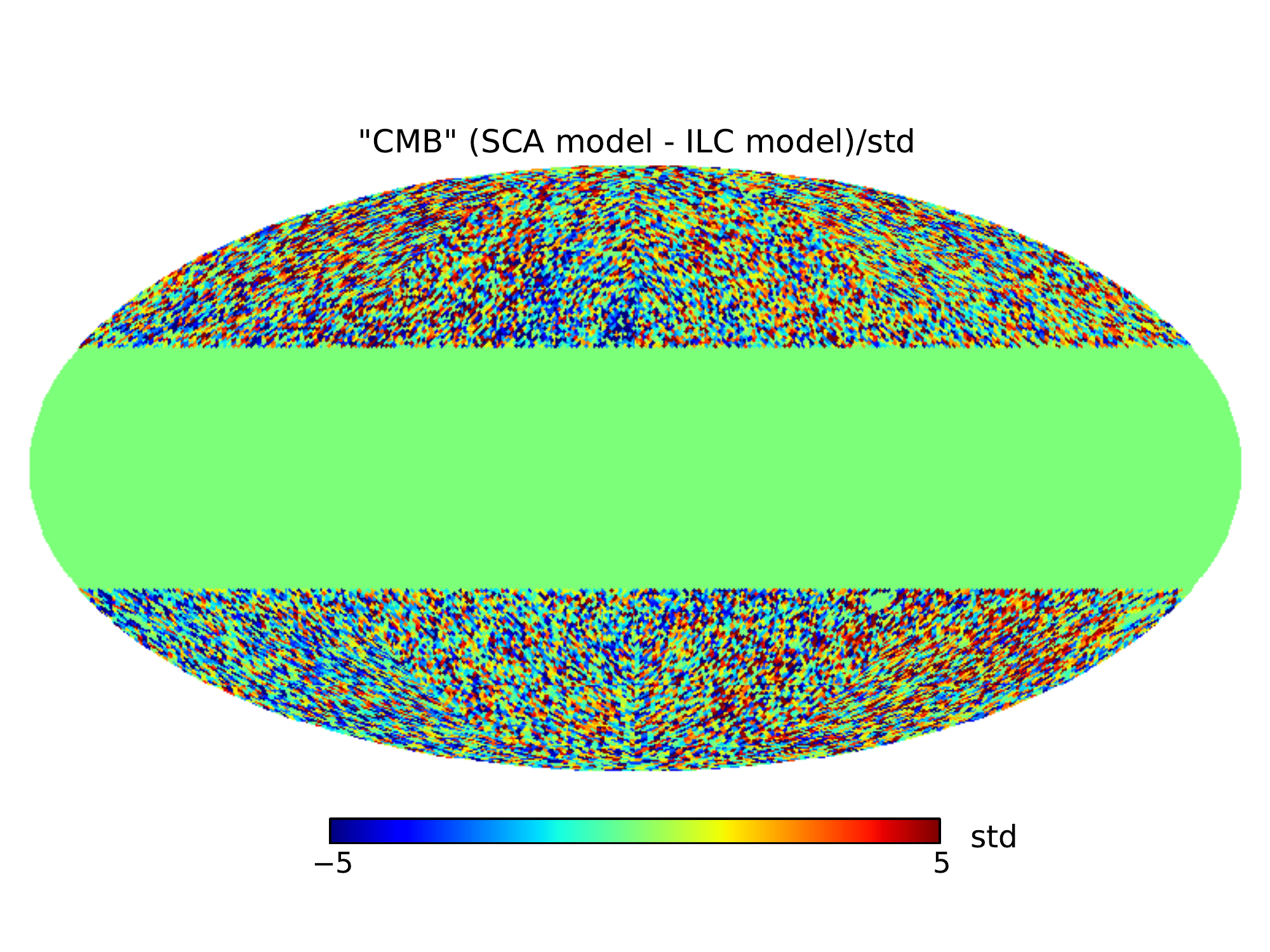, scale=\threepic}
\end{center}
\vspace{-8mm}
\noindent
\caption{\small 
Top: the SCA model of the CMB fluctuations.
Middle: the \WMAP ILC model of the CMB \citep{2011ApJS..192...15G}.
Bottom: the difference between the SCA model and the 
\WMAP ILC model of the CMB divided by the instrumental noise.
The SCA CMB model is very similar to the \WMAP CMB model,
but has a little higher random noise level.
}
\label{fig:CMB}
\vspace{1mm}
\end{figure}

\begin{figure}[t] 
\vspace{-5mm}
\begin{center}
\epsfig{figure = 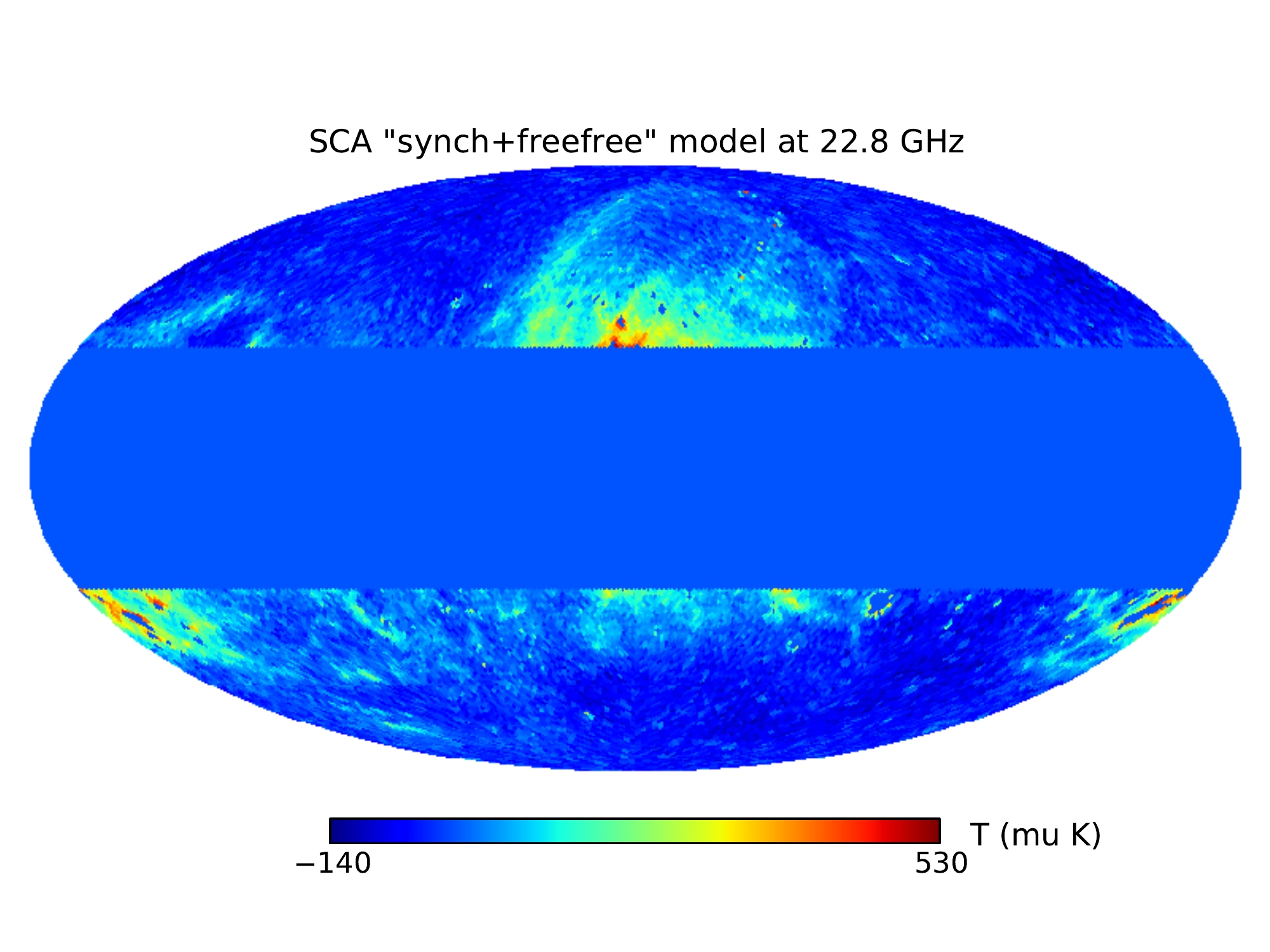, scale=\threepic} \\
\vspace{-5mm}
\epsfig{figure = 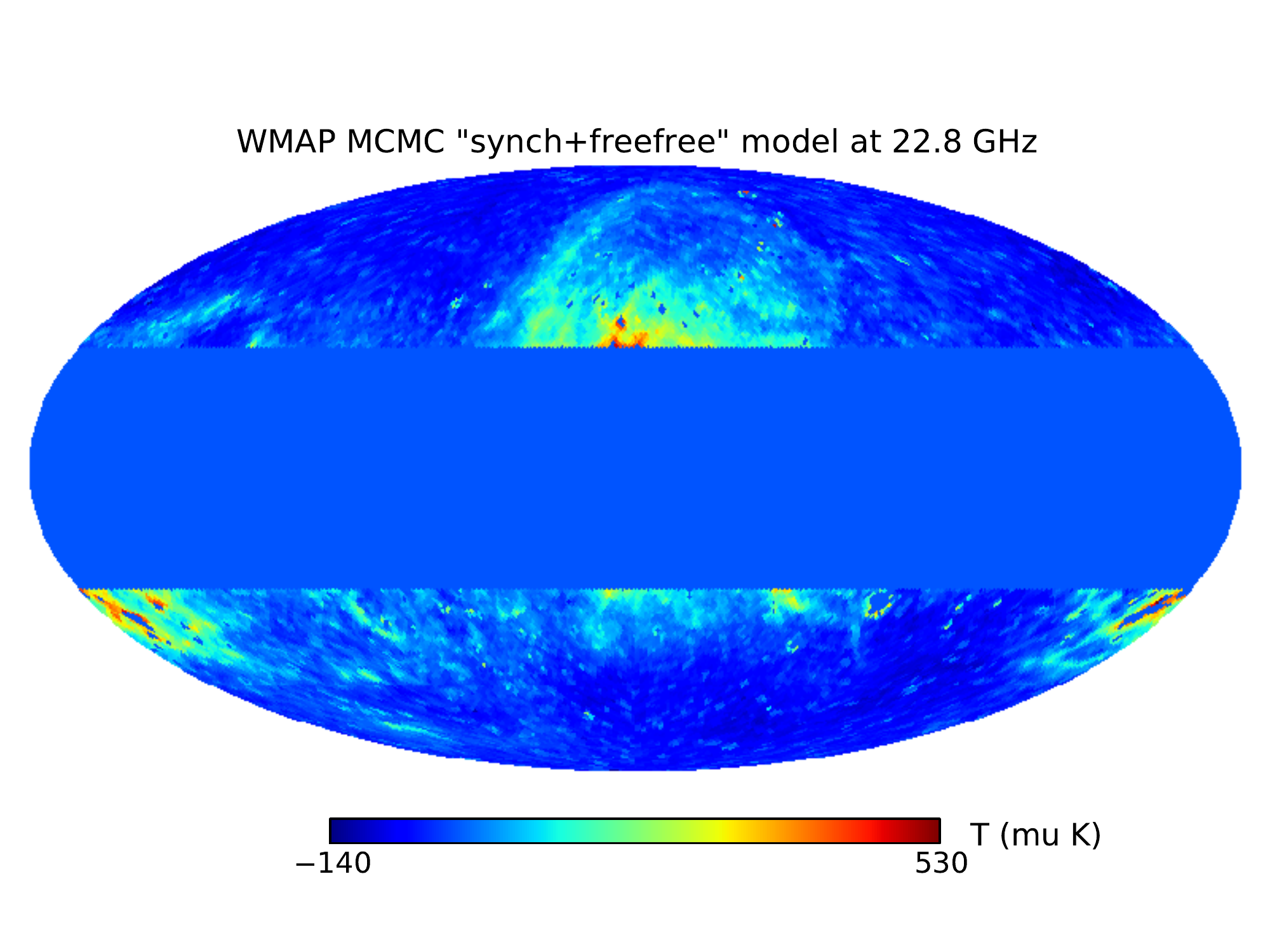, scale=\threepic} \\
\vspace{-5mm}
\epsfig{figure = 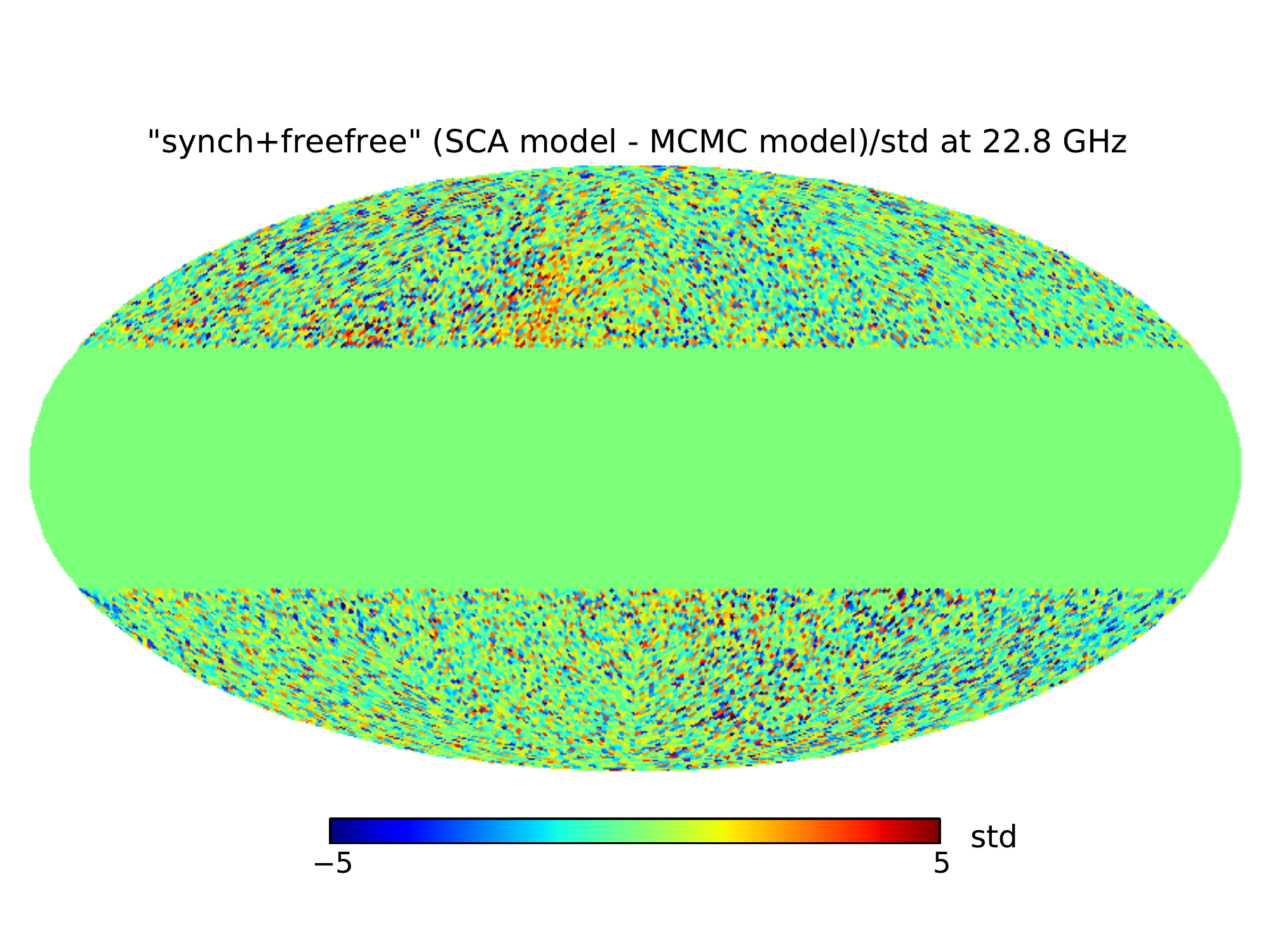, scale=\threepic}
\end{center}
\vspace{-8mm}
\noindent
\caption{\small 
Top: the SCA model of the synchrotron+free-free emission.
Middle: the \WMAP MCMC model of the synchrotron+free-free emission 
\citep{2011ApJS..192...15G}.
Bottom: the difference between the SCA model and the 
\WMAP MCMC model divided by the instrumental noise.
}
\label{fig:synch}
\vspace{1mm}
\end{figure}

\begin{figure}[t] 
\vspace{-5mm}
\begin{center}
\epsfig{figure = 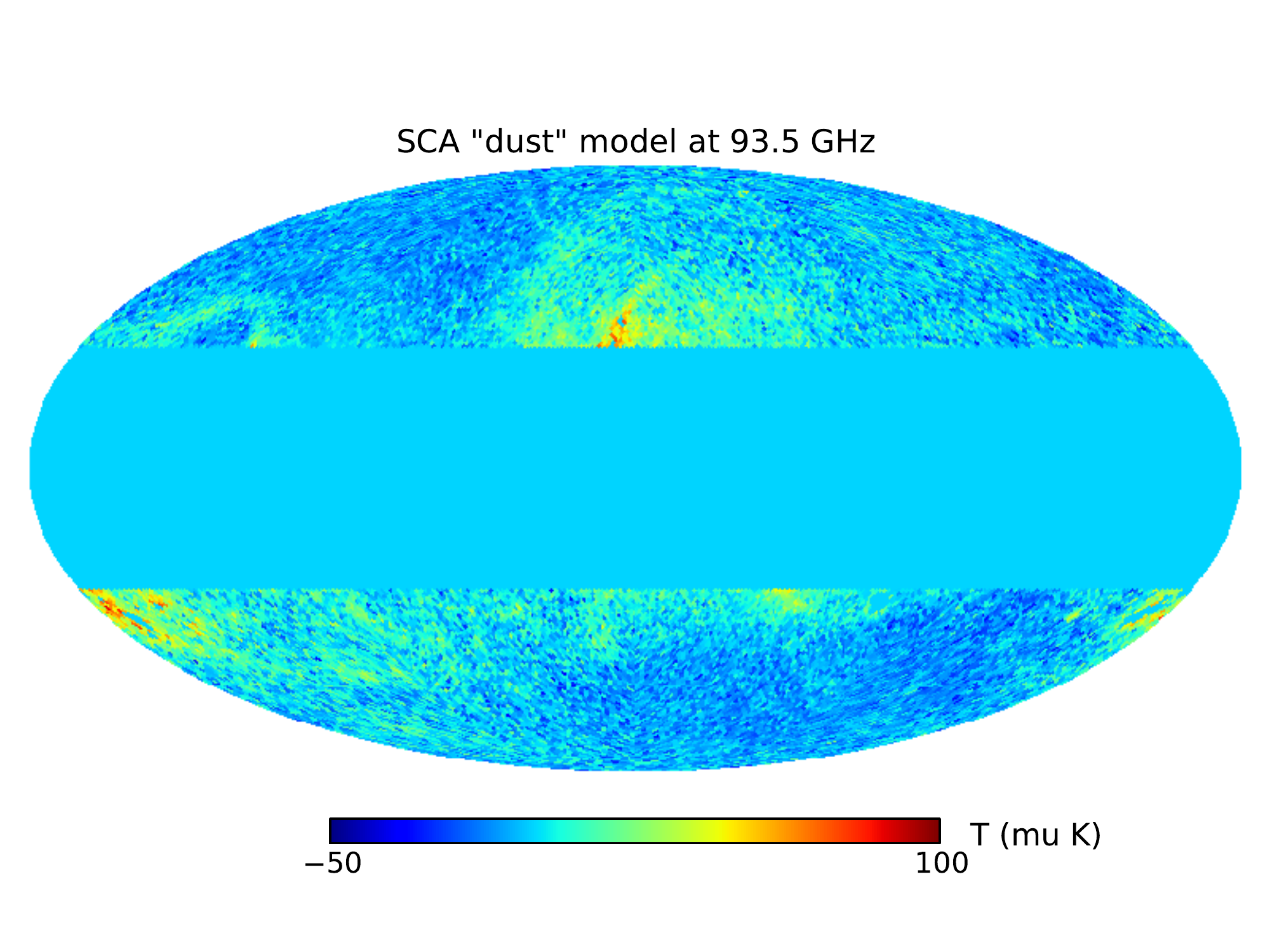, scale=\threepic} \\
\vspace{-5mm}
\epsfig{figure = 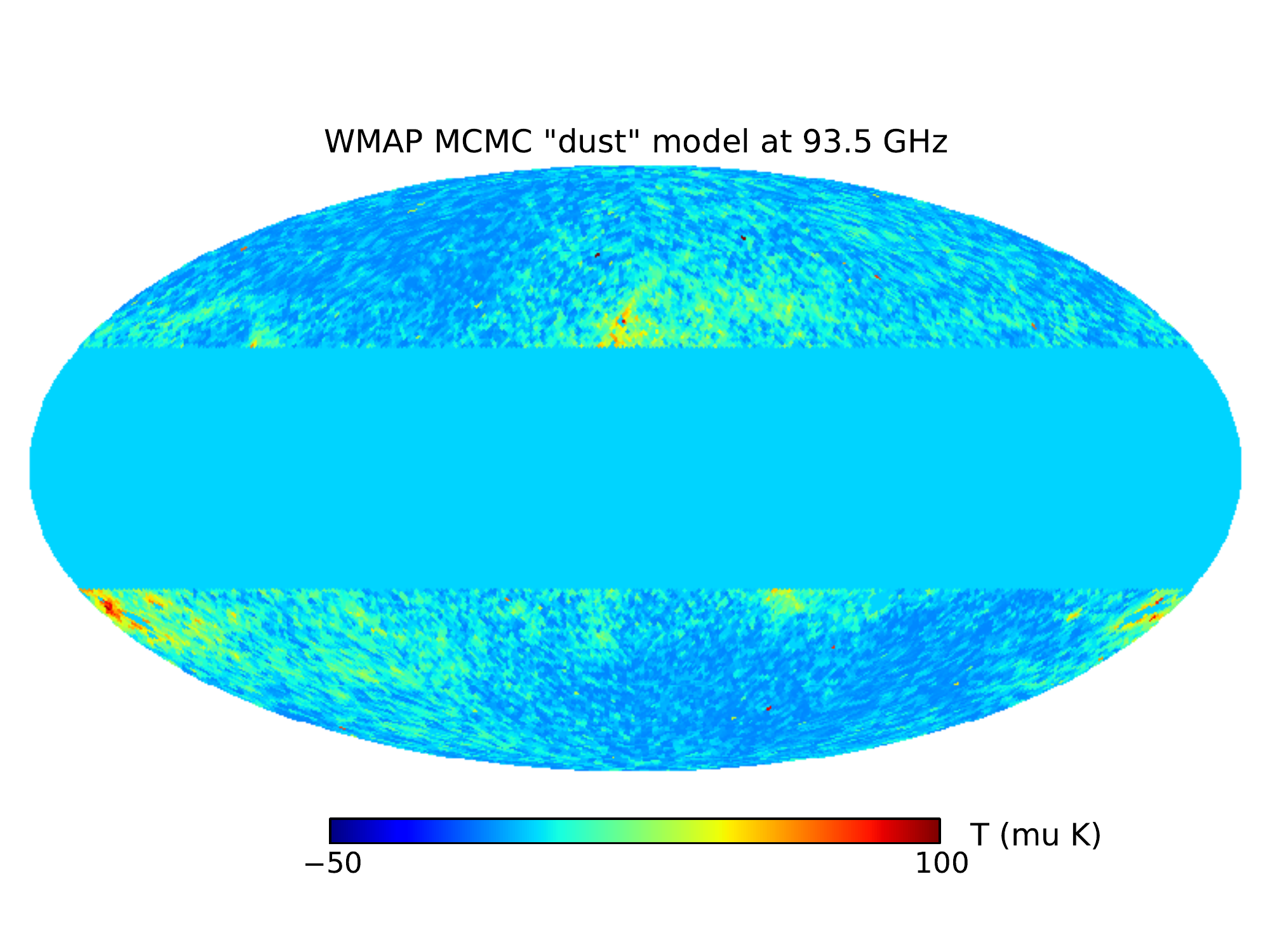, scale=\threepic} \\
\vspace{-5mm}
\epsfig{figure = 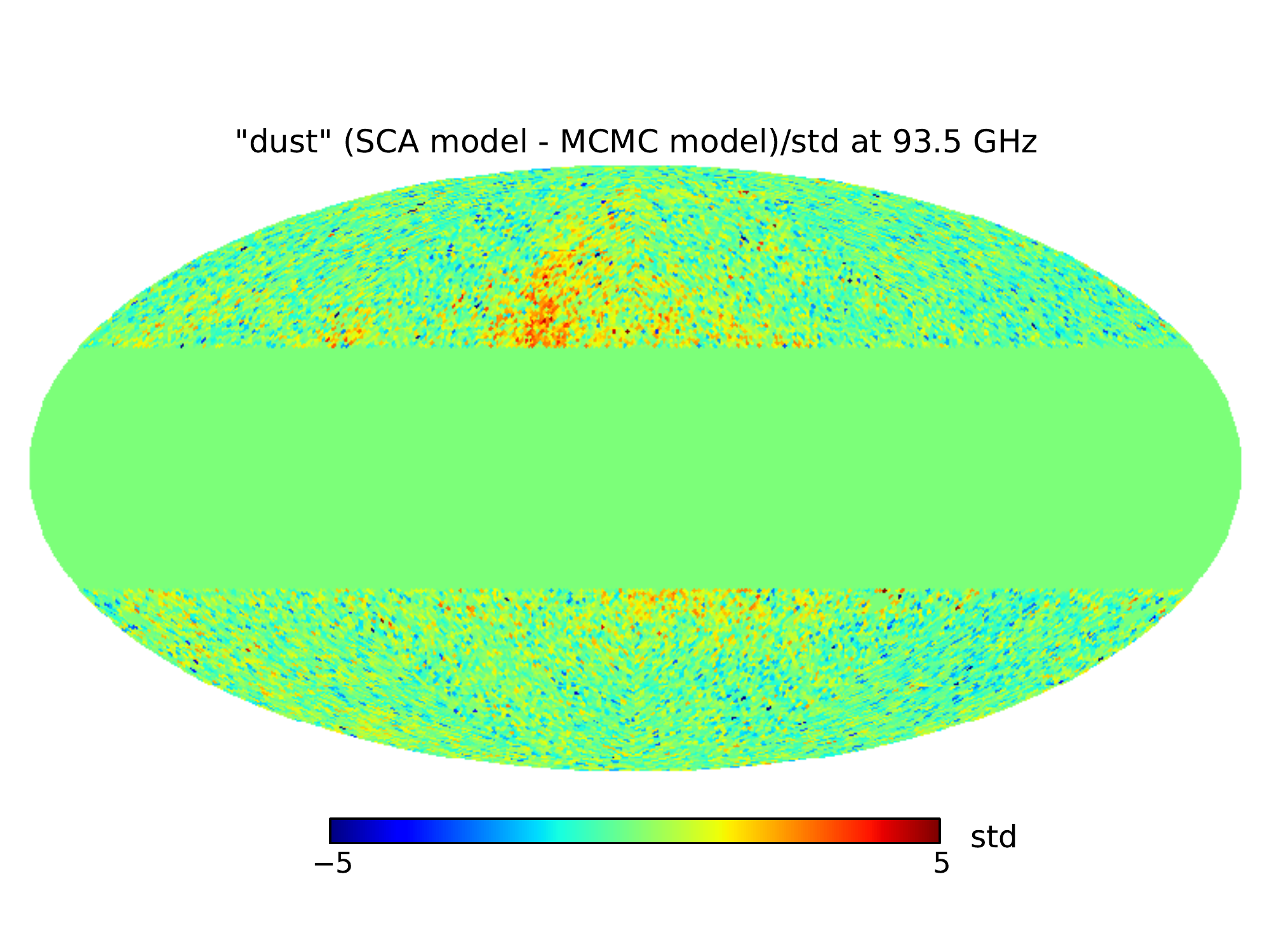, scale=\threepic}
\end{center}
\vspace{-8mm}
\noindent
\caption{\small 
Top: the SCA model of the thermal dust emission.
Middle: the \WMAP MCMC model of the thermal dust emission 
\citep{2011ApJS..192...15G}.
Bottom: the difference between the SCA model and the 
\WMAP MCMC model divided by the instrumental noise.
}
\label{fig:dust}
\vspace{1mm}
\end{figure}

\begin{figure}[t] 
\vspace{-5mm}
\begin{center}
\epsfig{figure = 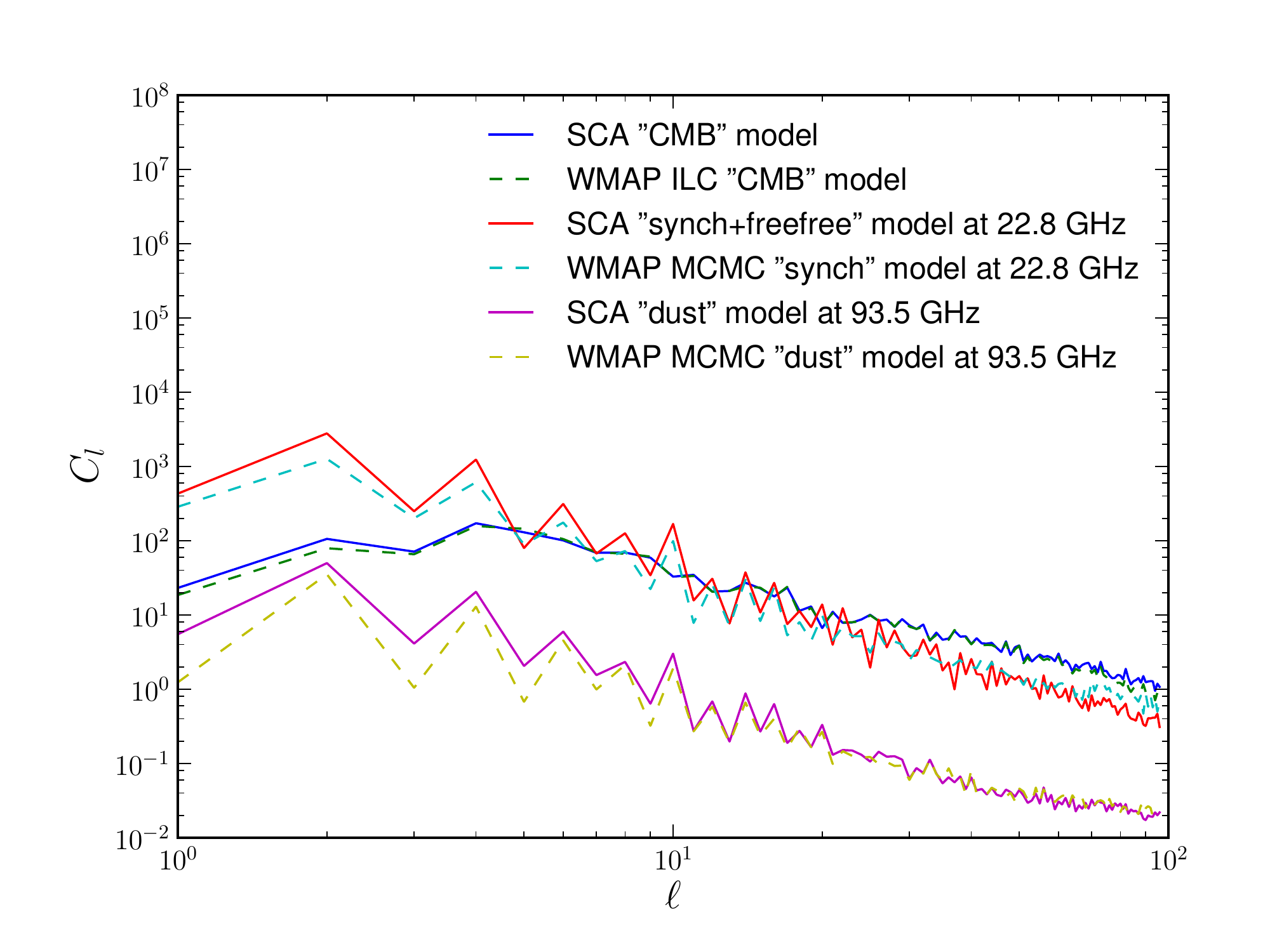, scale=\onepic}
\end{center}
\vspace{-8mm}
\noindent
\caption{\small 
Angular power spectra for the SCA models and the \WMAP models
presented in Figures \ref{fig:CMB}, \ref{fig:synch}, and \ref{fig:dust}.
The angular power spectra of the CMB are very similar to each other.
The SCA models of the synchrotron+free-free and the dust emission 
have a little more power at small $\ell$ than the \WMAP MCMC models.
Large up and down fluctuations of the foreground emission $C_{\ell}$'s
is due to the symmetry properties of the mask.
In calculating the $C_{\ell}$'s, we divide by the fraction of the sky that is unmasked.
}
\label{fig:cldiff}
\vspace{1mm}
\end{figure}

\begin{figure}[t] 
\vspace{-5mm}
\begin{center}
\epsfig{figure = 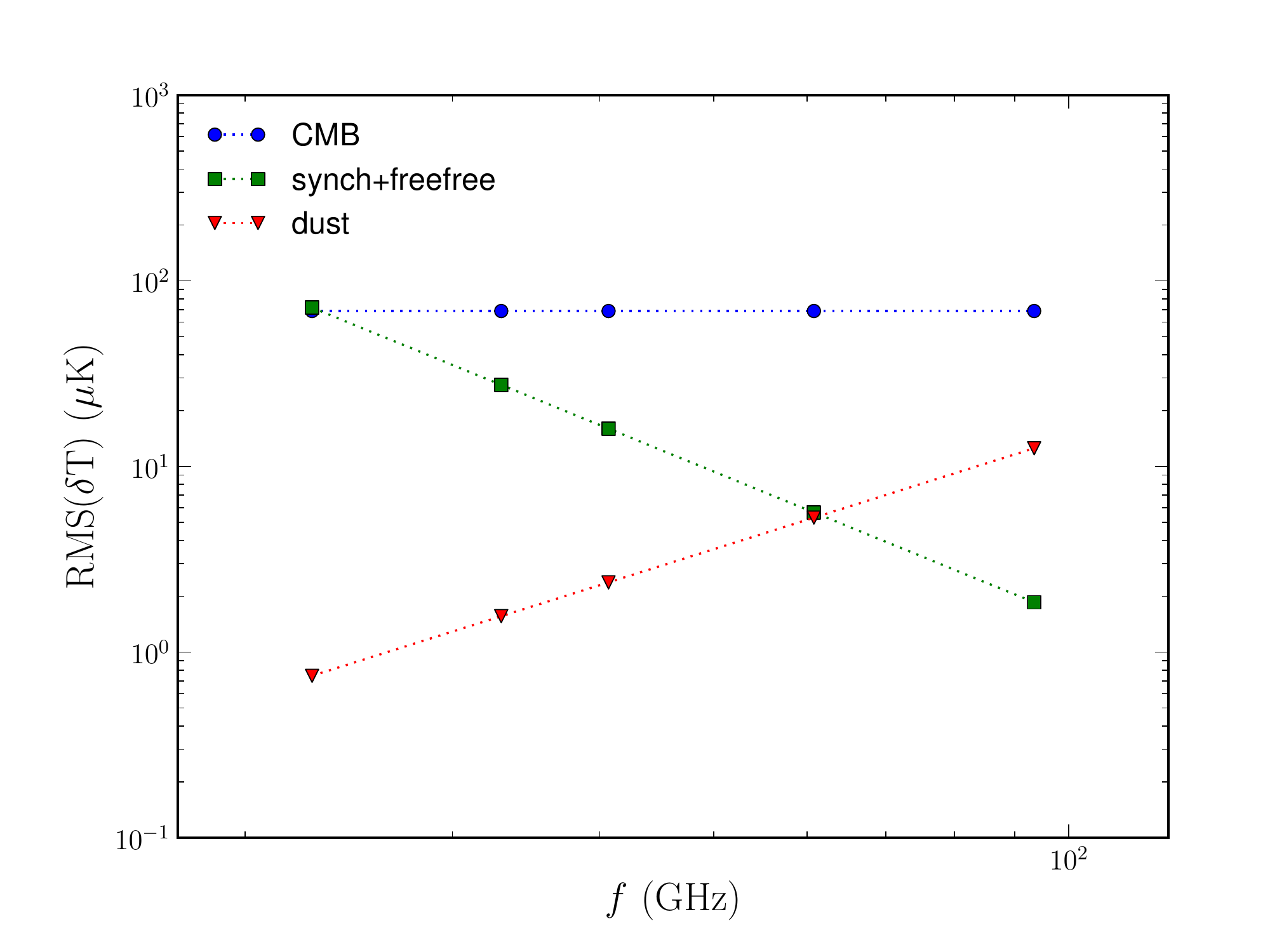, scale=\onepic}
\end{center}
\vspace{-8mm}
\noindent
\caption{\small 
Root mean squared of the temperature maps for the SCA models
(Figures \ref{fig:CMB}, \ref{fig:synch}, and \ref{fig:dust})
at different frequencies.
We assume power-law energy spectra.
The indices for the CMB and the thermal dust emission are fixed,
$n_{\rm CMB} = 0$ and $n_{\rm dust} = 2$.
The index of the third component and the normalizations
are found from fitting the models to the data in Section \ref{sec:example}.
The index for the third components is found to be $n_{\rm s+ff} = -2.6$.
This component is interpreted as a combination of the synchrotron and the free-free
emissions. 
}
\label{fig:rms}
\vspace{1mm}
\end{figure}

\bibliography{sca_papers}         


\end{document}